\documentclass[acmsmall,screen,nonacm]{acmart}

\usepackage{graphicx}
\usepackage{multirow}
\usepackage{multicol}
\usepackage{mdframed}
\usepackage{graphicx}
\usepackage{subcaption}
\usepackage{colortbl}
\usepackage{rotating}

\AtBeginDocument{%
  }

\begin{document}

\title{Co-Writing with AI, on Human Terms: Aligning Research with User Demands Across the Writing Process}

\author{Mohi Reza}
\orcid{0000-0001-9668-3384}
\affiliation{%
  \institution{University of Toronto}
  \city{Toronto}
  \state{Ontario}
  \country{Canada}}
\email{mohireza@cs.toronto.edu}
\author{Jeb Thomas-Mitchell}
\orcid{0009-0000-7403-2329}
\affiliation{%
  \institution{University of Toronto}
  \city{Toronto}
  \state{Ontario}
  \country{Canada}}
\email{mohireza@cs.toronto.edu}
\author{Peter Dushniku}
\orcid{0009-0002-3789-4629}
\affiliation{%
  \institution{University of Toronto}
  \city{Toronto}
  \state{Ontario}
  \country{Canada}}
\email{peter.dushniku@mail.utoronto.ca}
\author{Nathan Laundry}
\orcid{0000-0003-0846-2472}
\affiliation{%
  \institution{University of Toronto}
  \city{Toronto}
  \state{Ontario}
  \country{Canada}}
\email{nathan.laundry@mail.utoronto.ca}
\author{Joseph Jay Williams}
\orcid{0000-0002-9122-5242}
\affiliation{%
  \institution{University of Toronto}
  \streetaddress{40 St George St.}
  \city{Toronto}
  \state{Ontario}
  \country{Canada}
  \postcode{M5S 2E4}
}
\email{williams@cs.toronto.edu}
\author{Anastasia Kuzminykh}
\orcid{0000-0002-5941-4641}
\affiliation{%
  \institution{University of Toronto}
  \city{Toronto}
  \state{Ontario}
  \country{Canada}
}
\email{anastasia.kuzminykh@utoronto.ca}

\renewcommand{\shortauthors}{Mohi Reza et al.}
%% No italics, no superscripts
%% If needed use a foot or author note to identify equal contribution

\renewcommand{\sectionautorefname}{Section}
\renewcommand{\subsectionautorefname}{Subsection}
\renewcommand{\subsubsectionautorefname}{Subsubsection}
\renewcommand{\figureautorefname}{Figure}
\renewcommand{\tableautorefname}{Table}
\renewcommand{\equationautorefname}{Equation}

\begin{abstract}
   As generative AI tools like ChatGPT become integral to everyday writing, critical questions arise about how to preserve writers' sense of agency and ownership when using these tools. Yet, a systematic understanding of how AI assistance affects different aspects of the writing process---and how this shapes writers’ agency---remains underexplored. To address this gap, we conducted a systematic review of 109 HCI papers using the PRISMA approach. From this literature, we identify four overarching design strategies for AI writing support--\textit{structured guidance}, \textit{guided exploration}, \textit{active co-writing}, and \textit{critical feedback}--mapped across the four key cognitive processes in writing: \textit{planning}, \textit{translating}, \textit{reviewing}, and \textit{monitoring}. We complement this analysis with interviews of 15 writers across diverse domains. Our findings reveal that writers’ desired levels of AI intervention vary across the writing process: content-focused writers (e.g., academics) prioritize ownership during \textit{planning}, while form-focused writers (e.g., creatives) value control over \textit{translating} and \textit{reviewing}. Writers’ preferences are also shaped by contextual goals, values, and notions of originality and authorship. By examining when ownership matters, what writers want to own, and how AI interactions shape agency, we surface both alignment and gaps between research and user needs. Our findings offer actionable design guidance for developing human-centered writing tools for co-writing with AI, on human terms.
\end{abstract}

\begin{CCSXML}
<ccs2012>
   <concept>
       <concept_id>10002944.10011122.10002945</concept_id>
       <concept_desc>General and reference~Surveys and overviews</concept_desc>
       <concept_significance>500</concept_significance>
       </concept>
   <concept>
       <concept_id>10003120.10003130</concept_id>
       <concept_desc>Human-centered computing~Collaborative and social computing</concept_desc>
       <concept_significance>300</concept_significance>
       </concept>
   <concept>
       <concept_id>10003120.10003123</concept_id>
       <concept_desc>Human-centered computing~Interaction design</concept_desc>
       <concept_significance>300</concept_significance>
       </concept>
   <concept>
       <concept_id>10010147.10010178</concept_id>
       <concept_desc>Computing methodologies~Artificial intelligence</concept_desc>
       <concept_significance>300</concept_significance>
       </concept>
 </ccs2012>
\end{CCSXML}

\ccsdesc[500]{General and reference~Surveys and overviews}
\ccsdesc[300]{Human-centered computing~Collaborative and social computing}
\ccsdesc[300]{Human-centered computing~Interaction design}
\ccsdesc[300]{Computing methodologies~Artificial intelligence}

\keywords{AI-Assisted Writing, Human-Centered AI, Large Language Models}

\begin{teaserfigure} 
    \includegraphics[width=\textwidth]{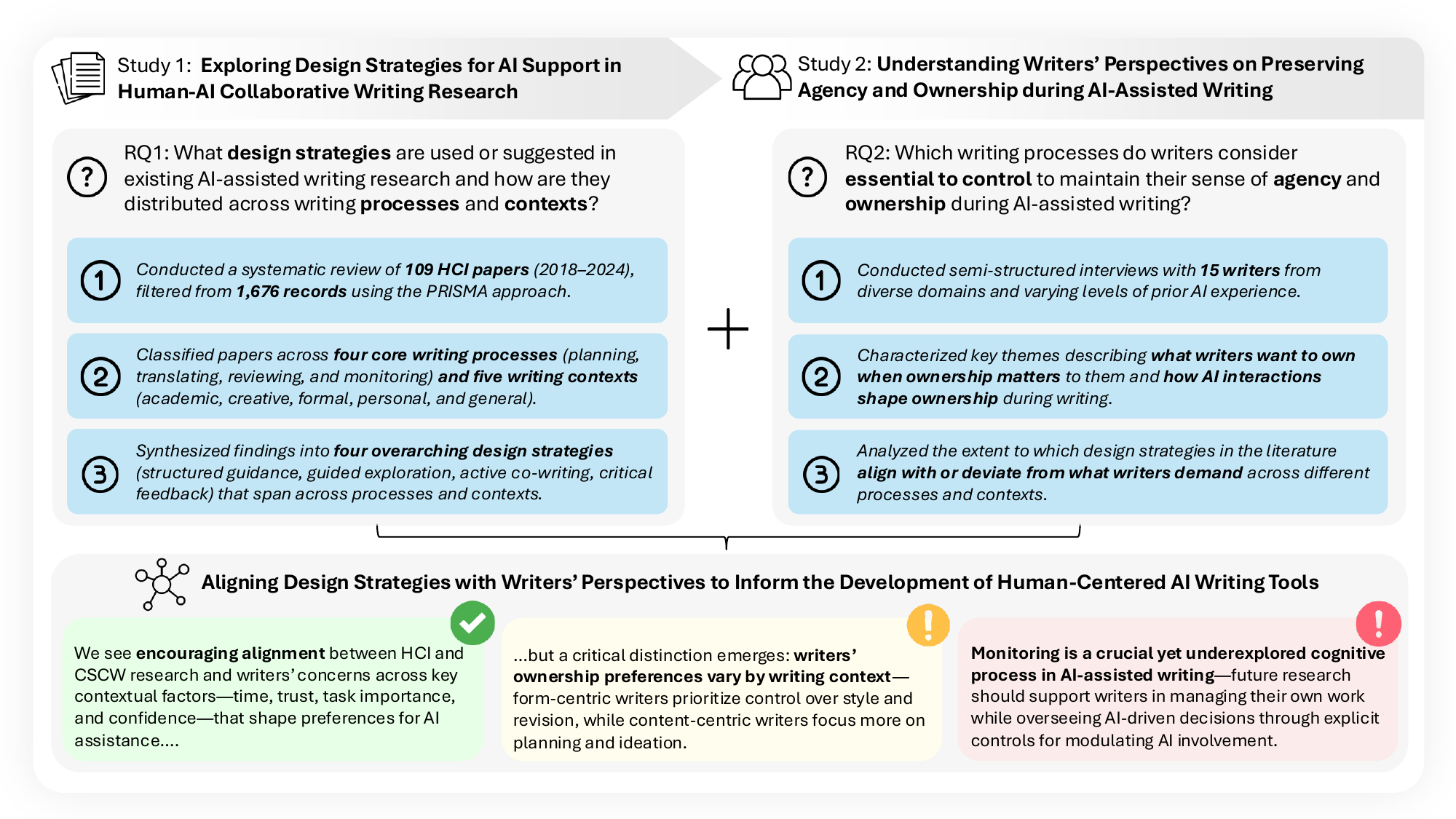} \caption{Research Overview: We present two interrelated qualitative studies exploring \textbf{how to design for human agency in Human-AI Collaborative writing}. (1) \textbf{Study 1}: A systematic review and thematic analysis of 109 papers (2018–2024), filtered from over 1,600 records in the Human-AI collaborative writing literature using the PRISMA methodology; (2) \textbf{Study 2}: A semi-structured interview study with 15 writers representing diverse writing genres, AI familiarity levels, and experience with generative AI tools.}
    \label{fig:overview}
\end{teaserfigure}

% \received{October 2024}
% \received[revised]{April 2025}
% \received[accepted]{August 2025}

\maketitle

\section{Introduction}

As Large Language Models (LLMs) grow more powerful and pervasive, AI tools like ChatGPT have become integral to the modern writing process. Computers are evolving from mere \textit{tools} to collaborative \textit{companions}, raising concerns about the encroachment of writers' co-creative boundaries \cite{biermann22}, eroding their agency and ownership over the writing process \cite{lee24, lee22}. At the same time, AI tools have proven to be remarkably helpful to writers, augmenting their capabilities across the composition gamut ranging from brainstorming and ideation \cite{shaer24, si2024can}, to editing and revision \cite{reza24, laban2024beyond}--and everything in-between \cite{mollick2024co}--making it impractical to abandon this technology altogether. 

One way to address the tension between ensuring human control and increasing automation \cite{shneiderman2022human} is to examine how existing and proposed AI-assisted writing systems in the Human-AI collaborative writing literature support distinct cognitive processes in writing \cite{flower81}, and whether that support encroaches on what writers consider to be central to preserving their sense of agency (i.e., their
perception of control and autonomy over the writing process), ownership (i.e., their feeling of
personal investment in and attribution of the final text), and task delegation (i.e., their choice about which writing subtasks to assign to AI, based on which cognitive processes they consider essential). Writers must maintain agency over the cognitive processes they value most through careful task delegation in order to preserve their sense of ownership over the final text. However, despite the unprecedented pace at which the AI-assisted writing research has grown over the past few years, there is currently no systematic understanding of what cognitive processes are being supported in numerous AI-assisted writing tools, what strategies are being used to offer that support, and how those strategies align with user perspectives across different forms of writing. Furthermore, these strategies have largely been evaluated from a usability and efficiency perspective \cite{binns2018s}, treating writing as a single task centered on optimization, with limited attention to preserving human agency through effective Human-AI collaboration \cite{amershi2019guidelines} and across its distinct cognitive processes \cite{flower81}.

In this work, we explore whether recent research on Human-AI collaborative writing aligns with user needs, shifting the focus away from output-oriented concerns like productivity toward human-centered considerations--particularly how to preserve writers’ sense of agency and ownership when collaborating with AI. To investigate this, we ask two research questions:

\begin{itemize}
     \item \textbf{RQ1:} What design strategies are used or suggested in existing AI-assisted writing research, particularly in terms of interaction models and the intended use of AI outputs, and how are these strategies distributed across writing processes and writing contexts?
    \item \textbf{RQ2:} Which cognitive processes do writers consider essential to control in order to maintain their sense of agency during AI-assisted writing, and how do user situations, writing contexts, and AI interaction types shape their perceptions of ownership?
\end{itemize}

To answer RQ1, we conducted a systematic review and meta-analysis of 1,676 papers in the Human-AI collaborative writing space from 2018-2024. We first analyze the systems developed or proposed in recent Human-AI collaborative writing literature, then classify them according to distinct thinking processes involved during composition, as outlined in Flower and Hayes cognitive process theory of writing \cite{flower81}. Then, to answer RQ2, we interviewed 15 writers across diverse domains, exploring how AI affects their sense of control and creative ownership across different writing processes. We then synthesized findings from both studies, revealing encouraging alignments as well as notable gaps between current AI writing support systems and writers’ expressed needs. \autoref{fig:overview} provides an overview of our research approach.

Our synthesis offers actionable guidance for designers, highlighting specific areas and methods of AI support that prioritize user agency. Rather than supporting all aspects of writing indiscriminately, our work helps focus design efforts on features that meaningfully preserve writers' sense of control and ownership. Grounded in these findings, our contributions to the CSCW Human-Centered AI community include:
\begin{itemize}
    \item The first comprehensive study on designing for human agency in AI-assisted writing, combining a systematic review of post-generative AI research with a user-centered analysis of how writers seek to preserve ownership and originality.
    
    \item A detailed characterization of four overarching design strategies for AI writing support, grounded in writers’ perspectives on when ownership matters, what they want to own, and how AI interactions shape that ownership.
    
    \item Actionable design guidance for CSCW and HCI researchers developing AI writing tools, including concrete recommendations for supporting writer agency across the cognitive processes of writing. Our work informs future systems that foreground meaningful Human-AI collaboration, rather than automation alone.
\end{itemize}

\section{Background \& Related Work}
This section provides the theoretical and empirical foundation for our study. We begin by justifying our selection of the Flower and Hayes cognitive process model as our analytical lens. We then review the evolution of AI-assisted writing systems, followed by a discussion of how these systems affect writers' sense of agency and ownership over the writing process.

\subsection{Theories on Writing Processes}
\label{sec: background_writingproc}
Early conceptualizations of writing processes by Rohman \cite{rohman65} focused on the temporal evolution of written documents. Rohman introduced a three-stage model emphasizing "pre-writing" – the preparatory phase where writers engage in thinking and analysis to discover patterns in their subject matter. This stage, followed by "writing" and "re-writing," was seen as essential for producing what Rohman termed "good writing" (i.e., text that makes original and insightful contributions). While groundbreaking, this linear approach would later be challenged by more dynamic models.

Flower and Hayes \cite{flower81} reconceptualized writing as a set of cognitive processes that writers deploy dynamically rather than in temporal stages. Their model identifies four primary processes: \textit{planning} (i.e., constructing internal representations of knowledge through generating ideas, organizing ideas, and goal-setting), \textit{translating} (i.e., transforming structured information into linear prose), \textit{reviewing} (i.e., evaluating and revising text according to established goals), and \textit{monitoring} (i.e., overseeing, regulating, and coordinating the writer’s cognitive activities, such as deciding when to shift between planning, translating, and reviewing, and ensuring alignment with writing goals). These processes operate within a task environment that includes the rhetorical problem and the emerging text, drawing upon the writer's long-term memory for topic knowledge and audience awareness. The processes form a hierarchical network where writers can move between processes at any time, or between high-level and local operational goals.

Nystrand \cite{nystrand89} expanded the theoretical landscape by incorporating social dimensions into writing process analysis. His framework emphasizes writing as a communicative event where meaning is actively constructed between writer and reader within discursive communities. Nystrand argued that skilled writers anticipate readers' expectations and manipulate their text to establish a temporarily-shared social reality. Hayes and Nash \cite{hayesnash96} detailed the cognitive architecture of planning, including planning by abstraction, analogy, and modeling. Kellogg \cite{kellogg87} explored the role of working memory in writing processes, while Hayes \cite{hayes96} expanded the original Flower and Hayes model to encompass social and physical environments, affect, and motivation. These contributions added depth to specific aspects of the writing process while building upon earlier foundational frameworks.

Our analysis employs the Flower and Hayes (1981) \cite{flower81} model as our primary theoretical lens for several reasons. First, it provides a comprehensive process model that describes writing behaviours. Second, its processes provide an analytical framework sufficient for examining the collaborative writing process between humans and AI systems. While Nystrand's model analyzes the social relationship between writer and reader, our research focuses on the collaborative interactions during writing, i.e. how humans and AI jointly engage in planning, translating, reviewing, and monitoring processes. The Flower and Hayes model allows us to examine how these cognitive processes are distributed and negotiated between human writers and AI systems during composition. Compared to other writing process theories, the Flower and Hayes model maintains an optimal balance between sophistication and analytical utility for our specific research context.

\subsection{AI-Assisted Writing}
Research on AI-assisted writing systems traces back to early implementations focused on creative writing support. Pre-transformer \cite{vaswani17} systems like Creative Help \cite{roemmele15} and Say Anything \cite{swanson12} utilized case-based reasoning and story repositories to generate context-aware sentence suggestions. Clark's \cite{clark18} work examining user experiences with AI writing prototypes revealed that while participants found AI collaboration satisfying, the resulting text quality did not surpass that of unaided human writers. These early systems laid the groundwork for understanding both the potential and limitations of AI writing assistance.

The emergence of transformer-based large language models in 2017 catalyzed research into AI writing assistance. In creative writing, researchers have developed systems supporting story writing \cite{yuan22}, playwriting \cite{mirowski23}, and character development \cite{qin24, schmitt21}. New systems support higher-level writing tasks such as prewriting \cite{wan24}, and generating perspective-specific feedback \cite{benharrak24}. Specialized creative applications have emerged for tasks including metaphor generation \cite{kimmetaphorian23}, collaborative storytelling \cite{nichols20}, and personal diary writing \cite{kimdiarymate24} as well as auxiliary creative tasks such as caption generation \cite{kariyawasam24}, title creation \cite{osone21}, and writing reflective summaries \cite{dang22}. Technical writing applications have focused on enhancing accessibility and supporting specialized writing tasks, including peer review \cite{wambsganss24}, literature reviews \cite{choe24}, and writing support for users with dyslexia or stuttering  \cite{ghai21, goodman22}.

User studies reveal complex dynamics in how writers interact with and perceive AI writing assistance. At a system interaction level, the design of AI suggestions significantly impacts user behaviour and output: sentence-level suggestions promote original content creation, while paragraph-level suggestions improve efficiency \cite{fu23}. Writers' engagement with AI assistance is also influenced by their personal values and goals. Writers show varying receptivity to AI support based on their confidence levels, demonstrating higher acceptance in areas where they lack expertise \cite{biermann22}, and their desires for support are closely tied to their perception of support actors and personal values \cite{gero23}. Moreover, this Human-AI writing relationship raises important concerns. Studies by Jakesch et al. \cite{jakesch23} and Poddar et al. \cite{poddar23} reveal that biased AI models can influence not only the resulting text but also users' own opinions. While users often value AI writing assistance highly, particularly for creative tasks \cite{li24}, professional writers note persistent challenges with AI systems' ability to maintain consistent style and voice \cite{ippolito22}. These findings highlight a central tension: as AI writing systems become more sophisticated, they must balance providing assistance while preserving authenticity and agency.

\subsection{Agency and Ownership in AI-Assisted Writing}

Recent research has examined how AI writing assistance affects users' sense of agency (i.e., their perception of control and autonomy over the writing process) and ownership (i.e., their feeling of personal investment in and attribution of the final text). Studies have shown that writers' sense of agency is significantly impacted by the level and type of AI intervention in the writing process. Robertson et al. \cite{robertson21} found that autocomplete suggestions could threaten users' autonomy. Similarly, Dhillon et al.'s \cite{dhillon24} research demonstrated that while next-paragraph suggestions improved writing quality, longer AI text completions decreased satisfaction by undermining writers' independence. This finding aligns with Draxler et al.'s work \cite{draxler24}, which showed that increased AI support corresponded with decreases in users' perceived control.

The relationship between AI assistance and text ownership is influenced by multiple factors, particularly professional context and writing purpose. Lee et al. \cite{lee22} identified a direct correlation between self-reported ownership and the proportion of user-written versus AI-generated text. Biermann et al. \cite{biermann22} found that storywriters who emphasized the expressive and emotional value of writing insisted on maintaining direct control over translation, viewing this control as essential to preserving their writerly identity and integrity. Gero et al.'s research \cite{gero23} revealed that the idea generation phase can particularly threaten ownership, with some writers considering the struggle with writer's block as integral to their writerly identity.

Several studies have identified factors that influence users' sense of agency and ownership in AI writing systems. Kobiella et al. \cite{kobiella24} found that participants who viewed AI as an enhancement tool rather than a replacement reported stronger feelings of accomplishment and ownership, while those who perceived their contributions as minimal experienced diminished ownership. Rezwana et al.'s \cite{rezwana23} work highlighted that ownership perceptions depend on both contribution levels and leadership in the writing process, suggesting that interaction designs that maximize user agency can enhance ownership. Through a thematic analysis of Human-AI co-creation discussion threads on reddit, Xu et al. \cite{xu24ownership} surfaced the themes of involvement, sense of infringement, and notion of legal agreements as most relevant in developing a sense of ownership. These findings indicate that maintaining user agency and ownership requires careful consideration of interaction design, user control mechanisms, and the balance between AI support and user autonomy.

These investigations have demonstrated the need for continued focus on users' senses of agency and ownership when writing with AI. Previous work on ownership in HCI, such as Kuzminykh \& Cauchard's framework \cite{kuzminykh20ownership}, has explored ways to develop a theoretical conceptualization of ownership in a traditional HCI context. However, there are currently no broad reviews of the AI-assisted writing research landscape that have evaluated HCI researchers' and system designers' strategies against users' needs for preserving their agency and ownership throughout their writing process.

\section{Study 1: Reviewing Writing Process Dimensions in the Literature}
\label{sec:study_1}
% % Introduction to study and connect to RQ1
To answer \textbf{RQ1: }``What design strategies are used or suggested in existing AI-assisted writing research, particularly in terms of interaction models and the intended use of AI outputs, and how are these strategies distributed across writing processes and writing contexts?'', we conducted a PRISMA systematic literature review on the ACM Digital Library database, and coded the resulting paper dataset, guided by the Flower \& Hayes Cognitive Process Theory of Writing \cite{flower81} and writing contexts, interfaces, and interactions enumerated by Lee et al. \cite{lee24}. We then performed thematic analysis \cite{braun2023doing} on the coded dataset in order to identify design strategies characterized by different interaction models, levels of AI support, and treatment of AI outputs. Finally, we coded the systems in our dataset by strategy in order to determine the distribution of the strategies across writing processes and contexts.

% % Brief preview of findings (1-2 sent.)
% Our analysis shows that current AI writing support systems employ diverse strategies to balance user agency with AI capabilities. These strategies differ primarily in how they position the AI, the intended outcomes of the Human-AI interaction, the design of the user interface, and how they manage the tension between efficient task completion and meaningful user engagement in the writing process.  

% Lay out the upcoming section/organization
\subsection{Methods}
We conducted a systematic literature review following PRISMA  guidelines \cite{prismapage} to identify and analyze research on AI writing support systems. PRISMA (\textbf{P}referred \textbf{R}eporting \textbf{I}tems for \textbf{S}ystematic \textbf{R}eviews and \textbf{M}eta-\textbf{A}nalyses) is a series of systematic review guidelines that are intended to improve the reporting and replicability of scientific literature reviews and meta-analyses. Our review focused on papers published between 2018-2024, corresponding to the emergence and widespread adoption of transformer-based language models \cite{vaswani17}. This period is marked by the explosion of AI research in HCI, visible in the publishing dates of papers in our dataset \autoref{fig:timeline}.

% \begin{figure}[H]
%   \centering
%   \includegraphics[width=0.5\linewidth]{figures/CSCW25-Timeline.pdf}
%   \caption{Distribution of papers by year, color-coded by HCI contribution type.}
%   \Description{Figure illustrating the }
%   \label{fig:timeline}
% \end{figure}

\subsubsection{Query Construction}
We developed our search query through an iterative process, beginning with a set of seed papers and expert knowledge in the field. We expanded our initial keyword list through multiple refinement cycles. The final search query combined writing-related terms with AI-related terms in order to capture as many potentially-relevant papers as possible: 
\vspace{0.5em} % Adjust the length as needed for desired spacing
\begin{mdframed}[backgroundcolor=gray!5, linecolor=black]
\begin{verbatim}
("writing" OR "writer" OR "write" OR "collaborative" OR "collaboration" 
OR "collaborate" OR "collaborating" OR "author" OR "authors" OR 
"creativity support" OR "co-creation" OR "co-writing") AND 
("AI" OR "language model" OR "artificial intelligence" OR "generative" 
OR "chatbot" OR "natural language processing" OR "NLP" OR "LLM" OR 
"digital assistant")
\end{verbatim}
\end{mdframed}

% (“writing” OR “writer” OR “write” OR “collaborative” OR “collaboration” OR “collaborate” OR “collaborating” OR “author” OR “authors” OR “creativity support” OR “co-creation” OR “co-writing”) AND (“AI” OR “language model” OR “artificial intelligence” OR “generative” OR “chatbot” OR “natural language processing” OR “NLP” OR “LLM” OR “digital assistant”) . 

\subsubsection{Exclusion Criteria}
We limited our review to peer-reviewed papers published in English, including journal papers, conference proceedings, and extended abstracts. We developed four primary exclusion criteria:

\begin{enumerate}
    \item[EC1.] Papers where AI interaction is not providing AI-assisted writing support, defined as AI writing with a user to create a natural language written artifact.
    \item[EC2.] Papers presenting purely technical, backend, or algorithmic contributions without user interaction.
    \item[EC3.] Papers focusing on non-natural language output formats (i.e., code or images exclusively).
    \item[EC4.] Papers that did not present a user study, artifact or system contribution, theory or conceptual framework, or systematic review.
\end{enumerate}

\subsubsection{Database Selection} 
To determine the optimal database for our review, we conducted a preliminary analysis across multiple digital libraries. We systematically sampled 100 papers from each of ACM Digital Library, IEEE Xplore, Taylor \& Francis, and Wiley by using a random sampling to select papers from search results for our query with 2018-2024 publication dates. We then applied our exclusion criteria to these papers' titles and abstracts. This preliminary assessment was designed to evaluate the concentration of relevant literature across databases. The ACM Digital Library yielded significantly more relevant results (11\%) compared to IEEE (2\%), Taylor \& Francis (3\%), and Wiley (3\%). Given this substantially higher concentration of relevant publications, we determined that the ACM Digital Library would provide the most comprehensive and targeted corpus of literature addressing our research questions.

\begin{figure}[H]
  \centering
  \begin{subfigure}[b]{0.48\textwidth}
    \includegraphics[width=\linewidth]{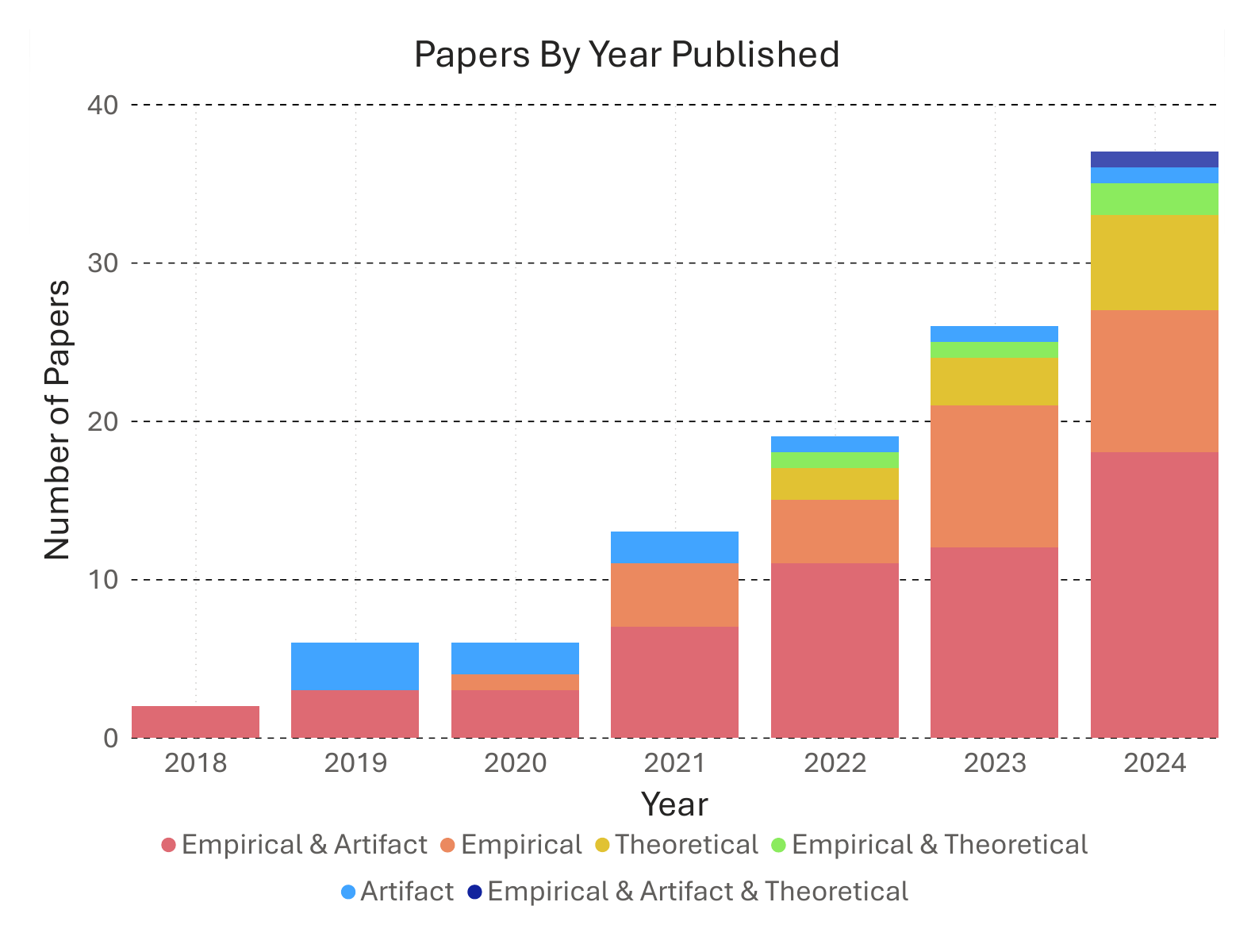}
    \caption{Distribution of papers by year, color-coded by HCI contribution type, highlighting the rapid growth of AI-assisted writing research following the introduction of transformer-based language models.}
    \label{fig:timeline}
  \end{subfigure}
  \hfill
  \begin{subfigure}[b]{0.48\textwidth}
    \includegraphics[width=\linewidth]{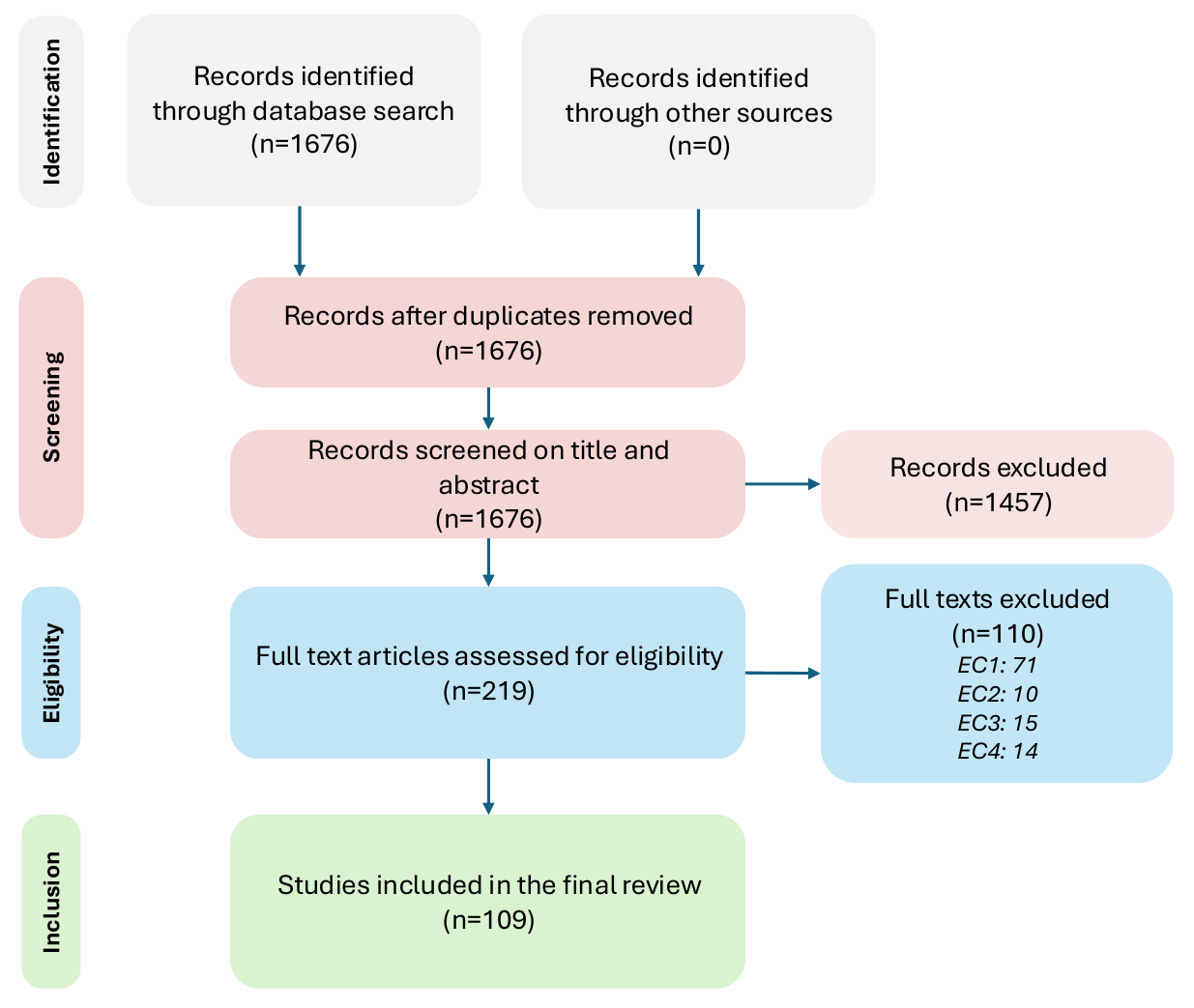}
    \caption{PRISMA flowchart illustrating the paper selection process for Study 1, including identification, screening, eligibility, and inclusion of 109 papers from an initial set of 1,676 records.}
    \label{fig:flowchart}
  \end{subfigure}
  \caption{Study overview: distribution of selected papers and paper selection process.}
  \label{fig:prisma_overview}
\end{figure}

\subsubsection{Screening Process} 
Our initial search yielded 1,676 papers. One researcher conducted the initial screening, applying our exclusion criteria to titles and abstracts, which identified 219 papers for full-text review. To ensure reliable coding, we conducted an inter-rater reliability test on a random sample of 25 papers from this set. Two researchers independently coded these papers based on the full text, achieving a Cohen's kappa of 0.84, indicating strong agreement\cite{mchugh12}. We resolved disagreements through discussion and consensus with a third researcher.

Following the confirmation of inter-rater reliability, we divided the remaining papers between two researchers for independent full-text review. This process resulted in the exclusion of 110 papers: 71 for not providing co-writing support (criterion 1), 10 for purely technical contributions (criterion 2), 15 for non-natural language output (criterion 3), and 14 for not meeting our paper type criteria (criterion 4). Our final dataset comprised 109 papers. \autoref{fig:flowchart} provides an overview of the paper selection process.

\subsubsection{Analysis}
We employed a codebook thematic analysis approach, developing our initial codes from Flower \& Hayes' cognitive process model  \cite{flower81} and Lee et al.'s design space framework \cite{lee24}. This complete codebook can be found under our supplementary materials. Following King \& Brooks \cite{king17} and Braun \& Clarke \cite{braun2023doing} we established our coding framework early in the process allowing us to inductively identify rich qualitative themes from the data.

% PRISMA flow diagram info
% - records identified through db search: 1676\\
% - records screened on Title/Abstract: 1676\\
% - records excluded on Title/Abstract: 1457\\
% - records assessed on full-text: 219\\
% - records excluded on full-text: 110\\
% - papers excluded by criterion:
%     - 1: 71
%     - 2: 10
%     - 3: 15
%     - 4: 14
% - final papers included: 109\\

\subsection{Study 1 Findings}

We first describe how writing support in the literature varies across five writing contexts, highlighting differences in the goals, users, and processes emphasized in each. We then introduce four overarching design strategies that characterize how systems support writers across these contexts, each with distinct implications for agency, task delegation, and interaction design.

\renewcommand{\subsubsectionautorefname}{Section}

\begin{table}[ht]
    \centering
    \renewcommand{\arraystretch}{1.3}
    \resizebox{\textwidth}{!}{%
    \begin{tabular}{|c|c|p{2cm}|p{2cm}|p{2cm}|p{2cm}|p{2cm}|}
    \cline{1-7}
    \multicolumn{2}{|c|}{\textbf{Cognitive Processes}} & \multicolumn{5}{c|}{\textbf{Writing Contexts}} \\ 
    \cline{3-7}
    \multicolumn{2}{|c|}{\textbf{}} & \textbf{Academic} & \textbf{Creative} & \textbf{Formal} & \textbf{Personal} & \textbf{General} \\ 
    \hline \hline
        \multirow{3}{*}{\textbf{Planning}} 
        & \textit{Generating} & \cite{neshaei24, singh24, choe24, ooz23, shibani24, shaer24} & \cite{osone21, kreminski20, lee22, clark18, schmitt21, gero19, peng23, ram21, wang22, chung22, yuan22, mirowski23, nichols20, shakeri21, kariyawasam24, qin24, ghajargar22, dang23, biermann22, inie23, gero23, singh23, tholander23, booten21, li24, wan24, kim24, hupont24} & \cite{kimmetaphorian23, gero22, arakawa23, fu23, bhat23, ding23, guo24} & \cite{arakawa23, lee22, shin22, wang22, zhang23, cremaschi23, dhillon24, cai24, janaka24, kimdiarymate24, li24} & \cite{chen19, gero19, lehmann22, buschek21, difede22, cremaschi23, bhat23, suh24, reza24} \\ \cline{2-7}
        
        & \textit{Organizing} & \cite{dang22, afrin21, resch19, taiye24, sun24, choe24, shibani24, shaer24, wang24} & \cite{kreminski20, schmitt21, wang22, chung22, yuan22, xu24, ghajargar22, kim23, biermann22, jahanbakhsh22, singh23, wan24} & \cite{kimmetaphorian23, kim23, ding23, jahanbakhsh22} & \cite{wambsganss20, shin22, wang22, zhang23, lin24, kimdiarymate24, jakesch23, poddar23} & \cite{goodman22, difede22, suh24, reza24, lin24} \\ \cline{2-7}
        
        & \textit{Goal-setting} & \cite{wambsganss21, resch19, taiye24, benharrak24, choe24} & \cite{kreminski20, qin24, yanardag21, biermann22, gero23, tholander23} & \cite{maiden19} & \cite{zhang23, cai24, janaka24, benharrak24} &  \\ \hline
        
        \multicolumn{2}{|c|}{\textbf{Translating}} & \cite{singh24, sun24, choi24, choe24, ooz23, goel23, liempowerment24, park24, shibani24, diloy24} & \cite{osone21, peng23, chung22, ghajargar22, dang23, yanardag21, biermann22, inie23, gero23, singh23, booten21, li24, kim24, hupont24} & \cite{kimmetaphorian23, gero22, fu23, bhat23, han24, liu22} & \cite{zhang23, lin24, cai24, janaka24, kimdiarymate24, jakesch23, liu22, draxler23, li24} & \cite{bhat23, suh24, lin24} \\ \hline
        
        \multirow{2}{*}{\textbf{Reviewing}} 
        & \textit{Evaluating} & \cite{wambsganss24, shen23, schmidt20, wambsganss21, ito23, pereira19, taiye24, benharrak24, singh24, choe24, allen18, liempowerment24, shibani24, singhbridging24, rawat24, wang24} & \cite{peng23, wang22, hoque24, xu24, yanardag21, gero23, liapis23, booten21, wu24, cheng24} & \cite{nouri23, robertson21, neyem24} & \cite{wambsganss24, wambsganss20, peng20, wang22, kim22, benharrak24, robertson21, cheng24} & \cite{cila22} \\ \cline{2-7}
        
        & \textit{Revising} & \cite{wambsganss24, shen23, ito23, pereira19, han23, benharrak24, choe24, huang20, ooz23, chang21, darvishi22} & \cite{yuan22, hoque24, booten21} &  & \cite{wambsganss24, wu19, peng20, zhang23, lin24, benharrak24} & \cite{ghai21, goodman22, difede22, reza24, lin24} \\ \hline

         \multicolumn{2}{|c|}{\textbf{Monitoring}} & \cite{wambsganss21, resch19} & \cite{liapis23} & \cite{arakawa23, sarrafzadeh21} & \cite{arakawa23} & \cite{cila22, muller22} \\ \hline
    \end{tabular}%
    }
    % \caption{Mapping citations to writing processes \cite{flower81} and contexts \cite{lee24}.}
    \caption{Mapping of cited papers to writing processes and writing contexts, based on our systematic review of AI-assisted writing literature. The table reveals uneven research attention across cognitive processes and contexts--for example, strong representation of Generating and Translating activities in Creative and Academic settings, and limited focus on Monitoring across all contexts.}

    \label{tab:table-1}
\end{table}

\subsubsection{Writing Context Characteristics}

% To analyze the patterns that are present in the existing literature for Human-AI co-writing, we categorized our data into five writing contexts. Each writing context possesses its own problem domain and user population, affecting the set of goals and requirements to operate in that context. We adapted our writing contexts from \cite{lee24}, with the notable alteration of combining Journalistic, Technical, and Professional writing into one category we call Formal writing, to capture the similarities these domains have in terms of writers' accomplishing well-defined tasks, with clear expectations as to the form of the writing they produce. In our analysis, we considered three cognitive processes from Flower \& Hayes: Planning (comprised of the Generating, Organizing, and Goal-setting sub-processes), Translating, and Reviewing (comprised of the Evaluating and Revising sub-processes) (see \autoref{sec: background_writingproc}). We did not include Monitoring in our analysis for two reasons: first, as discussed in \cite{nystrand89}, the workings of the monitor are left mostly unexplored in the original Flower \& Hayes framework, impeding our ability to define a code that we could apply to systems in our dataset; and second, we encountered no mention of this process by authors of papers in our dataset, whereas Planning, Translating, and Reviewing were commonly mentioned.

In the following section we characterize our dataset by the writing context where they offered support, based on contexts adapted from Lee et al. \cite{lee24}. We describe each writing context and how AI-assisted writing research supports each cognitive writing process across them; we also report how many papers\footnote{Note that the counts of papers do not add up to 109, the number of papers in our dataset. Although most papers only had a single context, a small number spanned multiple contexts.} were coded into each writing context, as shown in \autoref{tab:table-1}.

\begin{enumerate}
    \item \textbf{Academic (31 Papers).
} The Academic writing context includes papers that are focused on research, analysis, or educational use. Papers in this context include topics such as assistance with literature review \cite{choe24, wang24}, peer review \cite{neshaei24, sun24}, academic writing \cite{taiye24, singh24, ooz23, shibani24}, and essay writing \cite{dang22, wambsganss21, afrin21, resch19, benharrak24}. AI-assisted writing systems in this context are often focused on structured skill development for the users. Support for planning processes are typically intended to help users connect and structure their own ideas to accomplish complex tasks \cite{resch19, taiye24}. Translating support provides preliminary drafts or helps the user to restructure their ideas in a different form (e.g., point form to prose), but encourages the user to write and integrate ideas on their own \cite{sun24, singh24, choi24}. Reviewing support is delivered in qualitative form, such as suggestions or summaries \cite{shen23, benharrak24, singh24}. Systems do not revise the user's text directly, instead recommending improvements to prompt the user to revise the work themselves.

\item \textbf{Creative (37 Papers)}. Creative writing papers focus on artistic expressions and narrative-based texts. In the Creative writing context, topics include: story writing \cite{osone21, lee22, clark18, chung22, yuan22, ghajargar22, kim23, biermann22, singh23}, collaborative storytelling with AI \cite{kreminski20, nichols20, yanardag21}, including CSCW work on AI support for human collaborative storytelling \cite{shakeri21} and using dialects in creative writing \cite{wasi24diaframe}. Other assistance includes character creation \cite{schmitt21, qin24},  poetry \cite{booten21}, lyric generation \cite{ram21},  writing screenplays \cite{mirowski23}, and design fiction  \cite{tholander23}. AI also provides support with rhetorical or stylistic elements such as forming metaphors \cite{gero19} or learning vocabulary \cite{peng23}. A focus of researchers in this area is conducting empirical studies with writers to discover their writing strategies and requirements for support \cite{biermann22, inie23, gero23, li24, wan24, kim24}. Support for planning takes the form of generative ideation, usually presented as suggestions \cite{wang22, lee22}, although some systems have a more equal and collaborative storytelling focus that weaves the AI ideas into the story text \cite{osone21, kreminski20}. Support for translating often occurs simultaneously with support for generating ideas, creating narratives or creative elements that blend the user's prior text with new ideas from the AI \cite{ghajargar22, chung22}. Support for reviewing features a mix of quantitative and qualitative feedback, with a focus on the AI evaluating text and providing suggestions rather than revising it directly \cite{taiye24, choe24}.

\item \textbf{Formal (16 Papers)}. The Formal writing context represents professional, standardized modes of writing, characterized by structured forms, limited use of personal or emotional expression, and purpose-driven tasks that entail specific communicative goals. AI support from papers in this context is focused on topics like enhancing productivity \cite{arakawa23, maiden19}, writing business emails or reports \cite{fu23, liu22, neyem24}, reviews \cite{bhat23}, professional design problems \cite{ding23}, copywriting \cite{kim23}, document analysis \cite{jahanbakhsh22}, clinical use \cite{han24} and creating solutions to business problems \cite{guo24}. Planning support in this context is focused on extending and organizing the user's ideas, often through analogies and cross-domain reasoning \cite{kimmetaphorian23, gero22}. Translating support is focused on writing efficiency, enabling the AI to write in the same interaction location as the user, or to make suggestions that are integrated directly in the text \cite{fu23, bhat23}. Finally, reviewing support was limited, and focused on evaluation using quantitative feedback like readability metrics \cite{neyem24}, and visual feedback such as progress bars \cite{nouri23}.

\item \textbf{Personal (24 Papers)}. The Personal writing context concerns self-expression and sharing one's thoughts, feelings, and experiences. Compared to other contexts, it embraces informality, subjectivity, and authenticity.  Writing tasks in this context include non-academic opinion essay writing \cite{arakawa23, wambsganss20, lee22, zhang23, dhillon24, li24}, blog or social media posts \cite{shin22, cremaschi23, lin24, cai24, janaka24, benharrak24, jakesch23, poddar23}, personal messages \cite{kim2019love} and journaling \cite{kimdiarymate24}. AI support in this context is generally targeted at lay users, emphasizing ease-of-use. Planning and translating support are intermingled due to the frequent use of longer AI outputs that directly ideate and write for the user, though these are generally presented as suggestions in order to preserve the user's engagement with the text \cite{cai24, kimdiarymate24}. We also see transformation of user inputs between modalities like speech to text or visuals to text \cite{zhang23, lin24}, or between textual forms like keywords to prose \cite{kimdiarymate24}. Reviewing support is typically provided through quantitative feedback, with a focus on evaluation rather than direct revision \cite{wambsganss24, peng20}.

\item \textbf{General (15 Papers)}.
The General writing context contains systems that are presented for use in multiple contexts, or where the system design is not adapted to solving problems from a particular contextual domain. We also included systems that provide accessibility support in this context. Writing tasks include dyslexia support \cite{goodman22}, support for people with speech impediments \cite{ghai21}, writing both personal and professional emails \cite{chen19, lehmann22, buschek21}, and writing applications which are targeted at multiple contexts \cite{difede22, bhat23, reza24, lin24, suh24}. A recurrent theme in planning and translating support in this context was the provision of interfaces that enabled rapid iteration and organization of idea and text generations \cite{lin24, reza24, suh24}. Systems commonly provided suggestions for revisions which could be integrated directly into the writing area \cite{chen19, lehmann22, buschek21}, aiding efficiency and idea exploration in the text.
\end{enumerate}

\subsubsection{Strategies for AI-Assisted Writing Support in HCI Research}
\label{sec: study1_strategies}
Our analysis revealed four overarching design strategies for AI writing support that span cognitive processes and writing contexts. These strategies are distinguished by the AI's role, intended user behaviors, interaction outcomes, interface design, and usage of AI outputs. Systems can combine elements from multiple strategies based on their supported writing processes and contextual requirements. Each strategy offers varying support for writers' sense of agency, ownership, and task delegation preferences across different contexts. 

\begin{enumerate}
\item \textbf{S1: Structured Guidance}. This strategy represents a scaffolding approach where AI systems function as writing coaches or tutors, guiding users through document development while maintaining their autonomy and preserving agency. This strategy emphasizes active skill development through structured practice rather than passive reception of AI-generated content, typically requiring predefined writing tasks. The strategy comprises four key components. \textbf{Pattern Mapping} focuses on developing connections and pattern recognition within existing content rather than generating new ideas, with AI systems helping users locate patterns in their data and analyze potential suggestions. \textbf{Sequential Development} denotes an iterative approach through drafts and milestones, where the system guides users in adapting suggestions to build their writing capacity. \textbf{Scaffolded Feedback} delivers assessments through structured templates, combining quantitative metrics with clear evaluation frameworks, and encouraging the user to perform their own revisions. Finally, \textbf{Workspace Control} employs user interfaces that physically separate AI and user workspaces, ensuring users maintain control over textual changes while explicitly initiating support requests at each stage of the writing process. Revisions utilizes proposals which the user can reference, or analysis to help the user revise their text, which ensures the user still contributes to the text. This approach respects writers' need to maintain agency over ideation and organizing, thereby preserving their sense of ownership.

\item \textbf{S2: Guided Exploration}. This strategy positions AI systems as facilitators that enable users to actively explore and make connections within an idea space, with the AI functioning as both map-maker and guide. This strategy supports both well-defined and ill-defined writing tasks, emphasizing user engagement through iterative exploration and selection while maintaining creative control. It encompasses four main components. \textbf{Idea Navigation} implements a structured, self-directed approach that balances assistance with skill development, focusing particularly on interfaces which allow users to swap between generations to explore different approaches to their rhetorical problem. These systems enumerate the idea space using ideas generated by the AI. \textbf{Output Variation} denotes the provision of multiple types of output by the AI (i.e., narrative elements like plot and creative elements like dialogue), offering flexibility in AI generation. Systems directly replace user text in the writing area, enabling users to evaluate revisions in place, with the option of using the exploration interface to undo changes.\textbf{ Iterative Revision} utilizes the map of the idea space generated in exploration to both structure potential ideas and guide revision, facilitating an iterative model of exploration and refinement by the user.\textbf{ Proposal Integration} maintains user control by having the system present ideas and text generated by the AI as proposals. The focus on exploration offers the user flexibility in how to integrate generations into the artifact, with an emphasis on user-initiated AI output. This balances task delegation needs by allowing writers to maintain agency over idea selection while delegating generation, supporting their sense of ownership.

\item \textbf{S3: Active Co-Writing}. This strategy establishes AI systems as active writing partners, enabling a collaborative relationship where users selectively offload writing tasks while maintaining editorial control over the final output, though with potential implications for ownership. This strategy accommodates both well-defined and ill-defined tasks by supporting rapid iteration and efficient workflows. It consists of five primary components. \textbf{Direct Generation} involves direct generation of substantial content (i.e., full drafts or long text completions) intended for integration into the final artifact, encompassing both idea development and formal aspects of the text. \textbf{Content Conversion} preserves user ideas through various transformation types (i.e., foreign language translation, translating keywords to prose). The transformation retains the user's original meaning, utilizing the AI to deliver that meaning in new forms. \textbf{Efficiency Optimization }denotes prioritization of speed and usability through streamlined interactions, contrasting with skill-development approaches. These are often deployed in Professional contexts where productivity is paramount. \textbf{Turn-based Creation} denotes turn-based interactions through chat or collaborative storytelling, facilitating human and AI creative input with automatic integration of AI contributions into the final artifact.  Finally, \textbf{Result Ownership} maintains user control through suggestion selection and user-initiated AI output. However, unlike Proposal Integration, suggestions are integrated directly in the final text which may challenge writers' sense of agency by blurring task delegation boundaries.

\item \textbf{S4: Critical Feedback}. This strategy positions AI systems as editors and organizers, facilitating a user's reflective practice through structured feedback while maintaining a deliberate separation between the creation and analysis phases, supporting clear task delegation boundaries. This strategy requires well-defined tasks to enable evaluation and comprises four components. Unlike strategies that span the entire writing process, Critical Feedback represents a specialized approach where systems focus on reviewing and evaluation, maximizing analytical depth through structured assessments. \textbf{Qualitative Feedback} implements anthropomorphized or less-structured interactions that simulate tutoring scenarios through chat or natural language feedback. This method can provide revisions, but typically requires manual integration of suggestions by users. \textbf{Quantitative Analysis} provides structured assessments with a stronger focus on evaluation than revision, utilizing numerical or visual feedback. \textbf{Hybrid Evaluation} combines qualitative and quantitative approaches, using formal templates rather than conversational formats. This method offers a balance of revision and evaluation support that protects users' agency by requiring effort to integrate into the text. \textbf{Revision Guidance} connects analysis and organization by offering revision suggestions based on idea summaries and providing fine-grained tools for specific revision tasks (i.e., merging, rewriting, summarizing). \textbf{Analysis Separation} maintains user control through deliberate separation between AI output and user workspace, requiring user-initiation of AI output, and introducing friction by requiring manual integration of AI-proposed revisions. This design choice deliberately preserves the writer's agency over implementation decisions, reinforcing their ownership of the final text through strategic task delegation.
\end{enumerate}

\begin{table}
    \centering
    \includegraphics[width=1\linewidth]{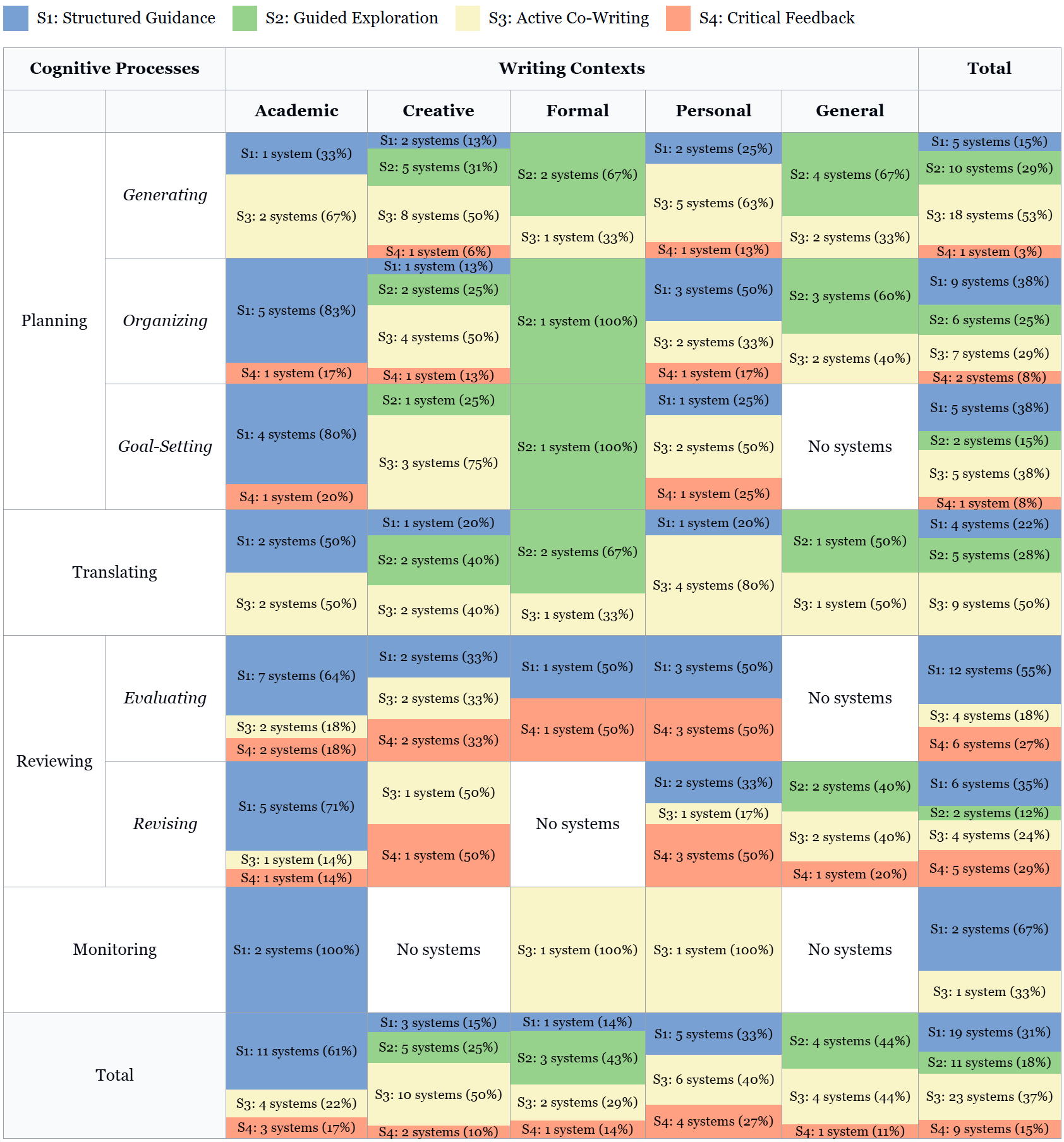}
    \caption{Distribution of systems by strategy across writing processes and contexts to show the prevalence of each design strategy in the literature dataset. Cell colouring is proportional to the prevalence of strategies deployed for systems in that cell, subject to a minimum height for readability. Systems could be coded to more than one process or context.}
    \label{fig:table2-proportional}
\end{table}

We applied our strategy framework to code the papers presenting system contributions (n=62) from our dataset to characterize the landscape of existing research systems. A single researcher assessed each system against the defining characteristics of each strategy. This coding revealed that S3 (Active Co-Writing) was the most commonly deployed approach (23 systems, 37.1\%), followed by S1 (Structured Guidance) (19 systems, 30.6\%), S2 (Guided Exploration) (11 systems, 17.7\%), and S4 (Critical Feedback) (9 systems, 14.5\%), with the full distribution shown in \autoref{fig:table2-proportional}. Clear patterns emerged across writing processes, with S1 dominating Evaluating (54.5\%) and S3 leading in Generating (52.9\%) and Translating (50\%) processes. Context-specific preferences were also evident, with Academic writing favoring S1 (61.1\%), Creative writing employing S3 (50\%), and Formal writing preferring S2 (42.9\%). Creative writing showed surprisingly high deployment of S3 systems, which are the most likely to threaten ownership. While the single-coder approach represents a limitation, this application of our framework highlights opportunities for more nuanced strategy implementation across cognitive processes.
\section{Study 2: Investigating AI's Influence on Ownership in Writing}
\label{sec:study_2}
% Our systematic review identified current trends and strategies in AI-assisted writing research and how existing systems support distinct cognitive processes involved in writing. 

% In order to answer \textbf{RQ3:} ``What cognitive processes do writers consider as essential to control in order to maintain their sense of ownership during AI-assisted writing, and how do user situations, writing contexts, and AI interaction types shape their sense of ownership?'' we conducted a interviews with 15 writers to understand their perspectives on which elements of the Flower \& Hayes Cognitive Process Theory of Writing \cite{flower81} are core to their sense of ownership and in what contexts. Then, to answer \textbf{RQ4:} ``To what extent do current AI writing support strategies align with or deviate from the support writers seek across different cognitive processes in order to preserve their sense of originality, ownership, and agency?'', we reflect on how the  strategies from the literature relate to user perspectives on how AI tools \textit{should} intervene, and how they influence writers' sense of ownership and control over different cognitive processes. Below, we describe the methodological details of the second study.

To answer \textbf{RQ2}: ``Which cognitive processes do writers consider essential to control in order to maintain their sense of agency during AI-assisted writing, and how do user situations, writing contexts, and AI interaction types shape their perceptions of ownership?'', we conducted interviews with 15 writers. 

\subsection{Methods}
We detail the methodological details of our second study, including participant recruitment, study procedures, and our approach to data analysis.

\subsubsection{Participants} 
We recruited 15 writers (8 women, 6 men, 1 did not specify; other gender options were offered) across two age groups: 5 participants aged 18–24 and 10 aged 25–34. Participants were based in North American and European cities and were recruited via social media and email invitations. They possessed diverse writing experience, including academic research papers (W11, W14), knowledge translation (W4), short stories (W2), poetry (W9), novels (W7, W13), essays (W1, W10), blogs (W4), screenplays for TV shows (W12), newspaper articles (W15), personal diaries (W5), internal project documentation (W6) and creative fiction (W3, W8). 11 participants reported having professional writing experience (i.e., when writing is paid or a core part of their occupation). Weekly time spent on writing varied, with 5 participants writing 1–4 hours, 5 writing 4–7 hours, 3 writing 7–10 hours, and 2 spending more than 15 hours per week.

Given our focus on AI-assisted writing, prior experience with AI writing tools was an inclusion criterion. All participants reported using ChatGPT, with Grammarly and Microsoft Copilot being the next most popular tools. Some advanced users also experimented with other LLMs and specialized AI writing tools, including Claude, LLaMA , and writing tools like Sudowrite. To ensure our findings were not biased toward users with a particular level of knowledge of generative AI, we recruited writers with varying generative AI expertise, ranging from slightly to extremely knowledgeable. The attached supplementary materials contain detailed information on writer profiles. 
\subsubsection{Procedure}

Each study session lasted between 60 and 90 minutes and was conducted online via recorded video calls by the lead author, allowing us to reach participants across multiple geographic locations. We introduced participants to the study and then asked them to complete a 5-minute pre-survey to provide consent and share demographic information, their writing experience, and AI usage. We informed participants of their right to withdraw from the study at any time and compensated each participant with CAN\$20 for their time. The institution’s research ethics board approved the study protocol.

\begin{figure*}
  \centering
  \includegraphics[width=\linewidth]{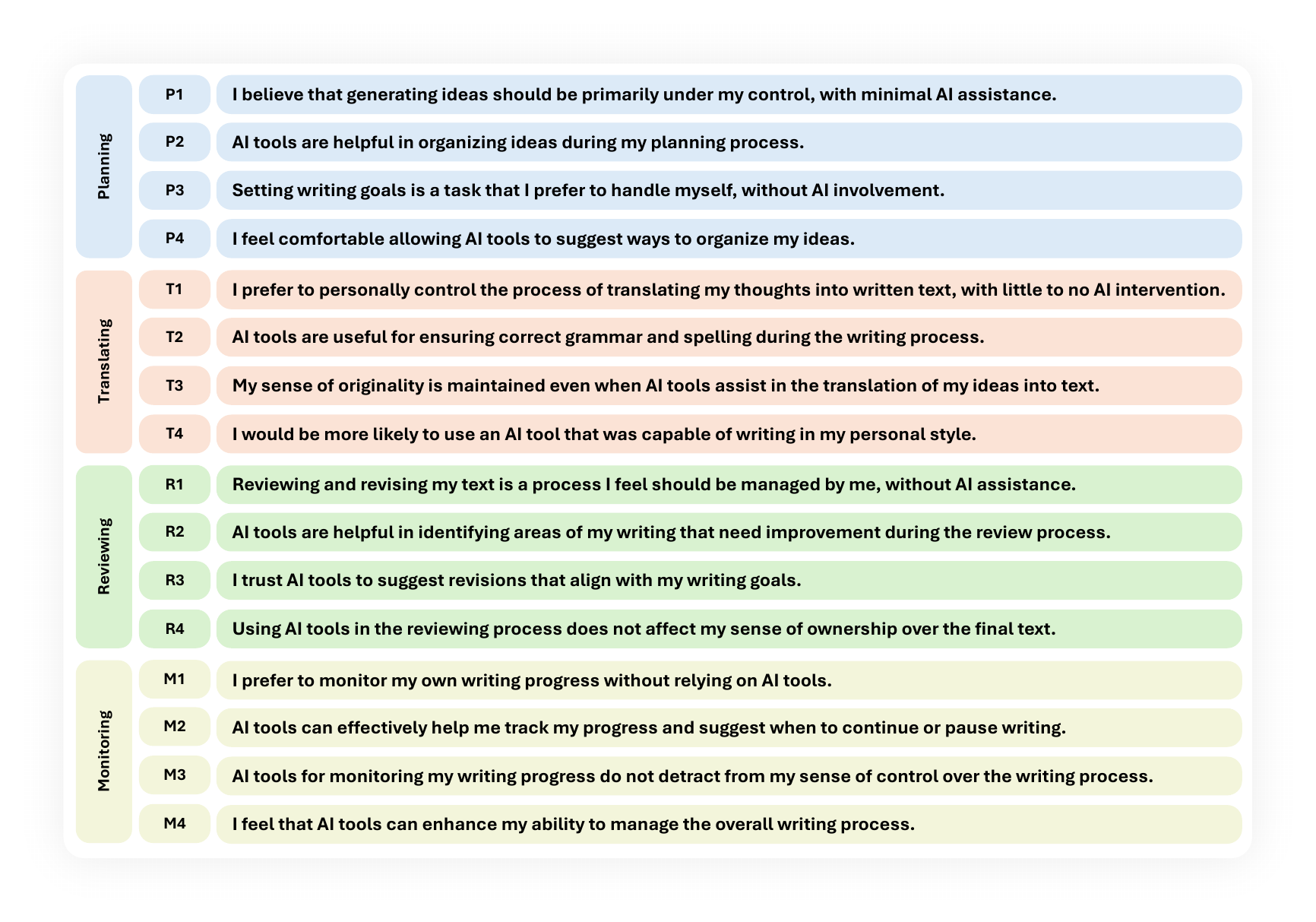}
  \caption{Likert-Scale Statements on User Perceptions of Cognitive Processes during Writing}
  \Description{Likert-Scale Statements on User Perceptions of Cognitive Processes}
  \label{fig:likert}
\end{figure*}

Following the pre-survey, we conducted a semi-structured interview in which participants described their writing background and experience with AI, and shared their perspectives on how AI influences their sense of ownership across each aspects of the writing process. We defined each process, to ensure writers could relate their practices to the processes. Finally, participants completed a 10 minute post-interview survey, reflecting on the discussion and rating 16 Likert-scale statements (4 for each process). This survey helped us gauge preferences for AI involvement in each element of the writing process. Further details on the post-interview survey Likert items are provided in \autoref{fig:likert}). The pre-interview survey and study protocol can be found under supplementary materials.

\subsubsection{Data Analysis}
The data included transcripts of the interview recordings and responses to pre- and post-interview surveys. To identify factors that influence writers' sense of ownership in the AI-assisted writing process, we conducted a reflexive thematic analysis \cite{braun2023doing} of transcripts through an inductive-deductive approach. Guided by the cognitive process theory of writing, we used the main writing processes—planning, translation, reviewing, and monitoring—as predefined codes to structure our interpretation, while also inductively identifying new patterns. The pre-survey data provided important context about each participant’s background in writing and prior experience with AI tools. The post-interview survey helped quantify attitudes toward AI across different cognitive processes. Scores for negatively worded items (P1, P3, T1, R1, and M1) were reversed (see \autoref{fig:bar}). Given the varied perspectives on ownership across writing elements, our goal was not to aggregate results into a single measure of ownership and agency but rather to examine distinct aspects of the writing process.

%It's okay to put findings in own section to avoid subsubsusbsection hell (gotcha)

\subsection{Study 2 Findings}

The post-interview survey responses are summarized in \autoref{fig:bar}. A detailed csv file of the responses can be found in the supplementary materials. Items are grouped into sets of four, labeled \textbf{P}1-4, \textbf{T}1-4, \textbf{R}1-4, and \textbf{M}1-4, corresponding to the four cognitive processes in the Flower and Hayes writing model: \textbf{P}lanning, \textbf{T}ranslation, \textbf{R}eviewing, and \textbf{M}onitoring. The item questions are detailed in \autoref{fig:likert}. The distribution of ratings reflects a range of perspectives on the extent to which writers want AI to intervene across different processes. We interpret this diversity through our thematic analysis, and share findings on how writers perceive and maintain a sense of ownership and agency over the writing process when working with AI. We group these insights under three primary themes, each highlighting a different dimension of the writers’ relationship to AI and their work:

\begin{enumerate}
    \item \textbf{When Ownership Matters}: This theme delineates the contextual factors–such as time constraints, level of trust in AI, task importance, and perceived competence–that shape writers’ decisions around how much control they want to retain and how much they are willing to delegate to an AI tool, even if it means their sense of ownership is encroached. Instead of assuming the desirability of ownership as an inherent or static prerequisite, this theme showcases the flexible role that human agency plays in AI-assisted writing and how it responds to situational factors. It also highlights situations where the risk of writers’ overreliance on AI is particularly prevalent. 

    \item \textbf{What Writers Want to Own}: This theme characterizes the aspects of the composition process and product from which writers derive their sense of ownership and prioritize as their primary contribution. We identify a central distinction between content and form: writers prioritize idea generation and planning as their primary contribution in content-oriented writing, where the purpose is primarily expository, while in form-oriented writing, where the focus is on style and voice, they emphasize the need to exercise more control during translation and revision to convey their unique expression.

    \item \textbf{How AI Interactions Shape Ownership}: This theme explores how interaction design impacts writers' senses of agency and ownership. We look at how different interface elements shape how writers feel when AI intervenes, such as the option to receive suggestions rather than direct edits, providing multiple suggestions, exercising final say, and UI affordances for enabling and disabling AI input. This theme highlights the critical role that Human-AI interaction design can play in maintaining writers’ sense of agency and ownership in AI-assisted workflows. 
\end{enumerate}
\begin{figure*}
  \centering
  \includegraphics[width=\linewidth]{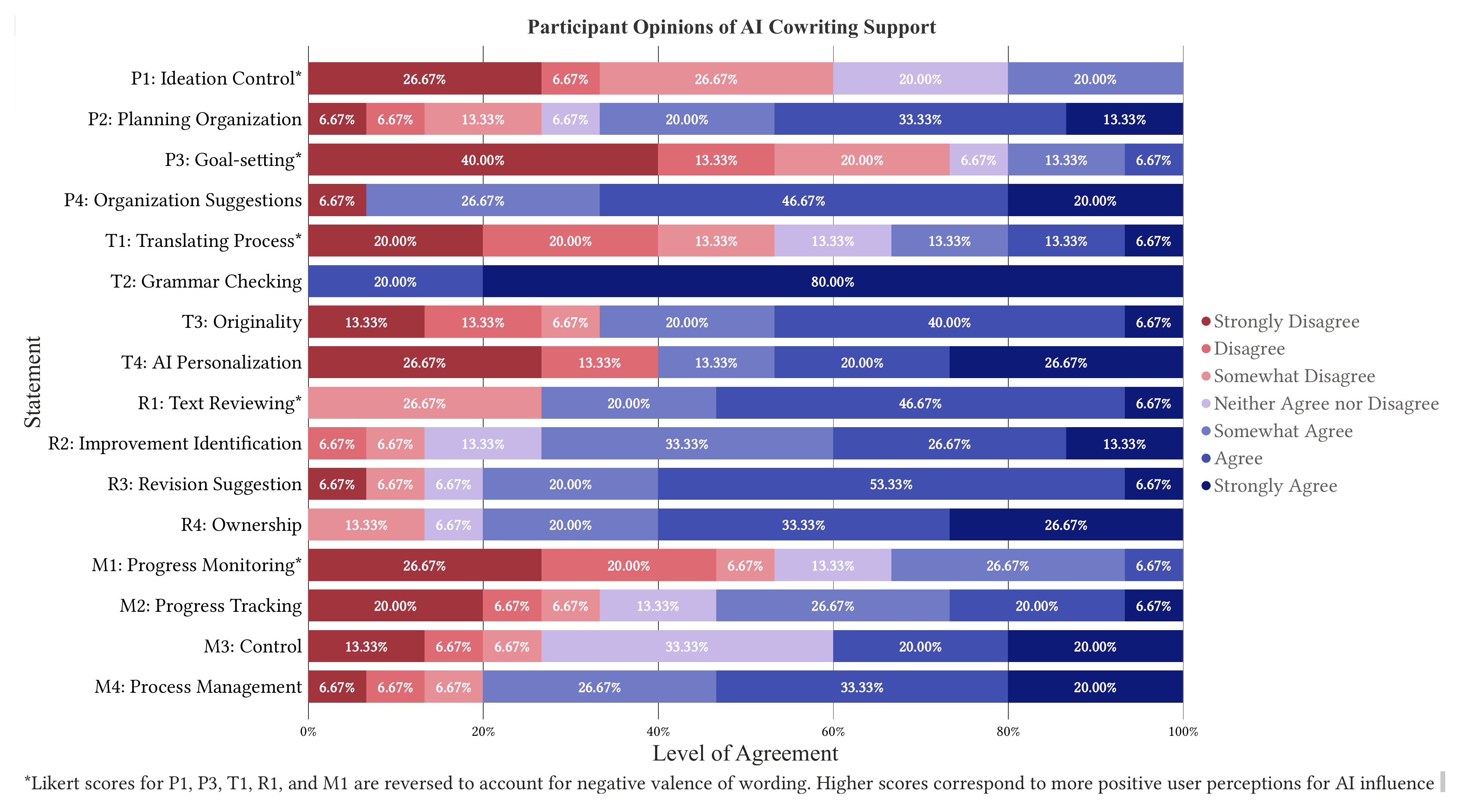}
  \caption{User perception likert-Scale items on writers' sense of ownership across cognitive processes \cite{flower81}. The distribution shows notable variation in the desirability of AI support across processes.}
  \Description{Figure illustrating the likert responses to survey.}
  \label{fig:bar}
\end{figure*}
Together, these three overarching themes offer a way to grapple with the complex interplay between writer agency, task demands, and AI functionality, helping AI system designers make sense of how ownership is negotiated and maintained in AI-assisted writing.

\subsubsection{When Ownership Matters}
\label{sec:when_matters}
We found four factors that influence how much control writers are willing to give to the AI and the extent to which they care about maintaining their sense of ownership in the first place -- \textbf{time} constraints, task \textbf{importance}, \textbf{confidence} in the writers’ own abilities, and \textbf{trust} in the AI’s capabilities.

\begin{enumerate}
    \item \textbf{Time:} A key reason writers are drawn to generative AI tools is efficiency. Therefore, in time-sensitive situations, writers are more willing to delegate tasks to the AI. W4 described ChatGPT as \textit{``a huge time saver''}, noting how \textit{``it sometimes helps when you're working on something super last minute, to have an AI look at it as well, and go through it in greater detail and precision''} than them. In addition to proofreading, writers are also more willing to delegate other aspects of the writing process. W11 shared how they used AI tools to transform rough bullets into polished writing. \textit{``There are also situations where I’m running short of time, and I will have a list of things I want to add…ordered in a reasonable way as I want them to appear in the writing. Then I will just ask ChatGPT to draft something based on the list.''} A similar point was echoed by W7, who described how they convert messy outlines into coherent text: \textit{``to save time, I will write out all bullet points myself that are really messy, and then have ChatGPT turn it into a letter.''}
    
    \item \textbf{Importance:} Writers de-prioritize ownership in low-stakes tasks, such as routine emails or straightforward professional communication, where clarity and efficiency are the primary goals. Such tasks tend to be perceived by writers as more functional than creative, making AI tools more acceptable for generating content without affecting their sense of ownership. The inverse is also true--when the stakes are high, writers become much less open to the idea of AI involvement, as captured by W2’s comment: \textit{``It depends how important this project is, because if it's very, very important to me, I would give [AI] less responsibility, almost to the point where it's just used as like something that I accept or reject, just like an editor who works for you, you can either accept or reject it or revise it… if it was not that important, like an email that I'm just kind of sending off. I would give it almost all the work, honestly.''} W15’s remark reveals how this choice to adjust the importance of ownership is deliberate, and not necessarily due to a lack of self-awareness -- \textit{``I use [AI] to finish my emails when Gmail tells me to finish with yours sincerely…I'm like, sure that's what I say. So for me, that's the sort of use of LLMs that I find quite pervasive in the background, and which I am definitely happy to use…I suppose there's an irony in pressing Tab to write `Yours sincerely’. You see, right? You're not being sincere.''}
    
    \item \textbf{Confidence:} For writers who feel less confident in specific language skills, AI tools can serve as a resource for checking for linguistic accuracy. Relying on AI for such help does not necessarily impact the writers’ sense of ownership, as W3 observes: \textit{``for grammar and spelling, those are inconsequential, right? I don’t associate that with the voice... It’s a menial task that can be taken care of by AI that doesn’t impact someone’s voice.''} This selective reliance on AI allows writers to focus their energy and sense of ownership on other parts of the work, while using AI to polish weaker areas. W3 elaborates on this intentional boundary: \textit{``I would want to make sure that anything that revolves around characters talking with one another, or whenever I write about the thoughts that the character is experiencing in their head, I’d want those to be my own work. But when it comes to describing a scene or a setting… that’s something that as a writer, I’m not that great at, and so it’s seeking help to make sure that my own work gets very polished.''}
    
    \item \textbf{Trust:} Just as confidence in their own abilities influenced writers' sense of ownership, their delegation choices were also shaped by their trust\footnote{The concept of trust and agency are interrelated, as they both influence users' decision-making abilities \cite{knittel2019most}.}—or lack thereof—in the AI's ability to deliver reliable output. This was particularly evident among writers who were confident in their own abilities but skeptical of AI. For example, W10 explained, \textit{``I don't particularly like the writing style. I don't trust it enough. There will always be a few nitpicks in any paragraph that I'll have with it. So in that way, I feel like I've been able to retain a total sense of ownership. I don't feel like it's influenced it any more than if I had someone read it and they said I liked it, or I didn't like it, or this part sucks.''} The ability to critically evaluate AI output helped writers like W10 maintain a sense of separation between looking at AI output and feeling like it inadvertently influenced them. W8 held reservations about the AI’s ability to gauge how humans would respond to a piece of writing -- \textit{``I just don't trust AI to judge whether something is understandable to a person or not, especially because of the variety of audiences I write for.''} Instead, they turned to humans for feedback, relying on colleagues and friends with different levels of expertise, from both within and outside their fields, to get varied perspectives on their work.
\end{enumerate}

% Define IBM Design Library colors
\definecolor{ibmblue}{HTML}{648FFF}
\definecolor{ibmpink}{HTML}{DC267F}
\definecolor{ibmyellow}{HTML}{FFB000}

% Create light versions for table cells (20% opacity)
\colorlet{ibmbluelight}{ibmblue!20}
\colorlet{ibmpinklight}{ibmpink!30}
\colorlet{ibmyellowlight}{ibmyellow!30}

\begin{table}[ht]
    \centering
    \caption{Delegation Strategies Based on Content and Form Contributions with Expanded Planning Categories}
    \renewcommand{\arraystretch}{1.3}
    \resizebox{\textwidth}{!}{%
    \begin{tabular}{|c|c|p{6.5cm}|p{6.5cm}|}
        \hline
        \multicolumn{2}{|c|}{\textbf{Cognitive Processes}} & \textbf{Content (e.g., Academic)} & \textbf{Form (e.g., Creative)} \\ \hline\hline
        \multirow{3}{*}{\textbf{Planning}} 
        & \textit{Generating} & \cellcolor{ibmpinklight} Strong ownership over novel ideas with minimal AI input. W9: \textit{``The core part of writing is ideas... even if ChatGPT helps organize, the ideas remain mine.''} & 
        \cellcolor{ibmyellowlight} AI-assisted ideation via prompts, but creativity retained by writer. W3: \textit{``I'd use [AI] for prompt creation... as a starter, then dive into writing.''} \\ \cline{2-4}
        
        & \textit{Organizing} & \cellcolor{ibmyellowlight} AI supports in structuring ideas, retaining ownership over logic flow. W9: \textit{``I let ChatGPT organize the ideas... but the logical flow is my own.''} & \cellcolor{ibmbluelight} AI-assisted outline creation for enhanced cohesion, writer's tone. W3: \textit{``Organizing can be AI-assisted if it doesn't alter my style.''} \\ \cline{2-4}
        
        & \textit{Goal-setting} & \cellcolor{ibmyellowlight} Writer defines key objectives and frameworks; AI used for background structure alignment. W13: \textit{``If I outline the goals clearly, ChatGPT can help format, but the primary direction remains mine.''} & \cellcolor{ibmyellowlight} Creative goals developed by writer; AI provides structure and adjustments. W7: \textit{``[AI] does help me understand the shifting goal posts that are moving around at a given moment.''} \\ \hline
        
        \multicolumn{2}{|p{2cm}|}{\vfill \textbf{Translating}} &  \cellcolor{ibmbluelight} AI used for drafting structured text; ownership tied to novel ideas, not genre standards. W7: \textit{``Academic writing feels less mine... I'm fine with delegating structure [to AI].''} & \cellcolor{ibmpinklight} AI assistance should be limited; primary voice retained through sentence-level decisions. W7: \textit{``Where I feel the most ownership is over sentences themselves.''} \\ \hline
        
        \multirow{2}{*}{\textbf{Reviewing}} 
        & \textit{Evaluating} & \cellcolor{ibmbluelight} Grammar and clarity editing delegated to AI for efficiency. W3: \textit{``For grammar, those are inconsequential... AI can handle it.''} & 
        \cellcolor{ibmpinklight} Limited AI assistance; primary voice retained through sentence-level decisions. W7: \textit{``Where I feel the most ownership is over sentences themselves.''} \\ \cline{2-4}
        
        & \textit{Revising} & \cellcolor{ibmbluelight} AI-intervention to refine writing clarity is welcome. W14: \textit{``I have a tendency to over-write, explaining things in a longer way; it can be much more concise. [AI can help] in that stage of the editing process.''} & 
        \cellcolor{ibmpinklight} Strong sense of novelty in breaking conventions for stylistic effect and unique voice. W7: \textit{``I break grammatical conventions for aesthetic effect... it's important for the voice, so I'm not interested in AI changing those choices.''} \\ \hline
    \end{tabular}%
    }
    \footnotesize
    \textbf{Legend:} 
    \colorbox{ibmpinklight}{Pink - Little Support} 
    \colorbox{ibmyellowlight}{Yellow - Moderate Support} 
    \colorbox{ibmbluelight}{Blue - Significant Support}
    \label{tab:categories}
\end{table}

\subsubsection{What Writers Want to Own}
\label{sec:what_matters}
In analyzing the areas from where writers draw their sense of ownership, we found a clear and recurring pattern: writers value ownership most strongly over components of the composition process they see as their primary contribution. Writers tend to be more open to delegating composition tasks to an AI for areas that are tangential to their perceived primary contribution. When task delegation is done this way, writers’ sense of autonomy and joy in creating something novel is maintained—or even enhanced in cases where the AI frees up their focus.

We identified two main types of contributions—content and form—each linked to specific cognitive processes. Content contributions involve generating ideas and setting goals, aligning with the planning process in the Flower and Hayes model \cite{flower81}. Form contributions focus on style, tone, and flow, aligning with translation and revision. This content-form distinction connects writing contexts with cognitive processes: academic and non-fiction writers prioritized content to convey ideas clearly, while fiction writers emphasized form, valuing their unique voice and style. Below, we explore writers' sense of ownership for these two types (see \autoref{tab:categories} for mapping):

\begin{enumerate}
    \item \textbf{Content:} When the purpose of a piece of writing is to convey pre-existing ideas or information with clarity–as is common in academic and non-fiction writing, the content itself becomes the primary contribution and the central focus of ownership for writers. In these writing contexts, writers derive a sense of ownership by engaging in cognitive processes involved in ideation and organization of ideas, making planning the dominant process that they seek to control. When the ideas are established, writers are open to using AI tools for translating ideas into clear language or reviewing their work to enhance clarity and polish. This is reflected in W9’s perspective on non-fiction: \textit{``I think the ideas are the core part of the writing. So if I'm giving ChatGPT the ideas that I want it to kind of organize, I think I still maintain that ownership of like, oh, those are my ideas, but it's enhancing my writing.''} 

    The desire to emphasize ownership over content rather than form is influenced by external constraints, such as word counts or stylistic conventions. W7 captures this sentiment when discussing the formulaic nature of research papers: \textit{``I feel less ownership of my writing in general, just because the rhetorical context in which I’m writing is so rigid and has such clear expectations... I feel like I did something interesting as part of the research project, but the write-up itself is just a write-up and nothing more. [I’m] fine to delegate the vast majority of that process to someone else.''} Since LLMs are trained on established standards, they can assist with refining text to meet these norms. 
    
    \item \textbf{Form:} In contrast, when writers have freedom over form, the idea of AI models intervening in the translation process becomes less appealing. W7, a professional literary fiction writer and novelist, explained how their sense of ownership lies in the \textit{``sentences themselves and how sentences are sculpted,"} emphasizing that this sentence-level decision-making is \textit{``what sets me apart as a writer.''} For them, style, rhythm, and structure are the personal touches they are not willing to delegate to AI or any other external influence. They further noted that while ideas and themes can feel culturally shared, the form in which those ideas are presented is where the writer’s individuality comes through: \textit{``Where I feel the most ownership as a literary author is over sentences themselves and how sentences are sculpted. So that's where I'm least willing to secede to anyone else, including an AI, because I consider kind of like my, my sentence-level decisions in large part. But what sets me apart as a writer... Whereas my ideas I think of as less oftentimes I use ideas which strike me as things which are kind of out in the ether culturally already, or it's not like each scene or particular decision I make conceptually is really distinct and different.''}
\end{enumerate}

Even if the language models were trained to mirror an author’s personal style, form-oriented writers could find this prospect wholly unappealing–\textit{``I would feel a little violated. I think for me, personal style is so signature to me, like to who I am, like, I don't want AI to be like, training itself on me and then trying to emulate me. ''}, remarked W12 also a professional writer. This sentiment underscores how, for writers whose sense of ownership is rooted in form and personal style, attempts to design systems that mimic their unique stylistic choices can clash with core values.

What about hobbyists? Turning to W3, a hobbyist fantasy fiction writer, who uses AI to handle parts of the ideation phase so they can “jump straight into writing.” They describe using AI to generate writing prompts and background settings, allowing them to focus on what they consider the core writing process: \textit{“I would occasionally use it for [writing] prompt creation. So just to kind of give [me] a little bit of a starter, just to have some kind of setting to work with so that I don’t need to spend a lot of time with the story building, the world building, at least the initial world building, and can just jump straight into writing.”} Here, W3 also uses language that separates the ideation process from what they see as core ``writing.'' Their phrasing indicates a view of ideation as a necessary setup that can be delegated to AI, while the actual crafting of sentences is where they feel personal investment and ownership.

\subsubsection{How AI Interactions Shape Ownership}
\label{sec:interactions}
The third and final theme explores how different types of Human-AI interactions impact writers’ sense of agency and ownership, as writers \textit{monitor} and make decisions on the various cognitive processes during writing. Interface features such as the ability to choose between AI-generated suggestions, maintain final decision-making power, and toggling AI assistance on and off allow writers to retain autonomy over their work. We find a common theme across these interactions: writers feel a stronger sense of ownership when they perceive themselves as having substantial control over the AI’s contributions, and that interaction design can shape these perceptions. We will illustrate this via four feature concepts that preserve ownership: \textit{AI Suggestions}, maintaining \textit{Final Say}, \textit{Global AI Toggles} and \textit{Local AI Toggles}:

\begin{enumerate}
\item\textbf{Suggestions}
Participants consistently shared that receiving AI \textit{suggestions} that they could accept, reject, or modify, was a non-negotiable aspect of preserving their sense of agency. This preference is underscored by the interaction mode where AI directly inserts or overwrites text within the users' writing space, which writers felt encroaches on their sense of ownership. W7 explained that suggestions maintained their sense of ownership by framing AI as an \textit{optional} aid rather than a co-author: \textit{``If it's sort of making suggestions, then it would not change my sense of ownership over the text, because I'd still feel like that's just sort of this pop-up window. But if it was inserting values in a more direct way, I think I would probably feel like I was losing some ownership.''} \\
    
\item\textbf{Final Say} W3 highlighted  the importance of having the \textit{final say} over AI-generated content: \textit{``I know that at the end of the day, if I ask it for help, it's not like, it's not a final say per se, right? It's not that I'm resigning my writing to the chatbot... if those suggestions turn out to be helpful, then I can continue with them, or I can set them aside as I see fit. So ultimately, I’m always in control.''}\\

W15 further described this decision-making as a ``negotiation'' with the AI, framing ownership as an iterative process of consciously selecting and refining suggestions: \textit{``I find it has to be a negotiation. I think, like, you see what the thing is suggesting, you think about that, and then you decide to take it on board. And I feel like that moment of decision and conscious interpolation of what it's suggesting... that's where the sense of ownership is not taken from you.''}\\

These examples indicate how writers preserve ownership by positioning the AI as an helper rather than a primary co-author. For primary contributions writers used the AI as source for inspiration, not substance. In less critical tasks, writers were open to using AI content selectively, to enhance efficiency without compromising ownership. But regardless of the stakes, they always wanted to have the final say. \\ 
    
\item\textbf{Global AI Toggle to Maintain Flow State:} Writers wanted the option to toggle AI suggestions on and off to minimize distractions. This is apparent in W7's description of their frame of mind during fiction-writing:        \textit{ ``In fiction writing, I really get in the zone, which is important to me so much that I like to block out even just sort of my background, my desktop, just everything... if it's at a moment where I'm editing anyway and sort of moving things around, yeah, I mean, especially if I had sort of like an intuition already... I would be happy to hear any and all suggestions from anyone, including an AI.''}\\
    
The ability to enter a “zone” or flow state reinforces writers’ ownership, as they feel more connected to the work without interference. Similarly, W1 emphasized the need for flexibility to open and close AI assistance as needed: \textit{``If there's like a feature where I can open for a suggestion, like a little separate tab on the right side of my screen, and I can always open and close it... when I'm like, really focused... I don’t have to care about what AI keeps suggesting, so as long as the user has that flexibility, it’s okay to keep focused.''}\\

\item\textbf{Local AI Toggle for Intentional Rule-Breaking:} Sometimes, instead of completely turning AI off, more advanced writers like W7 wanted fine-grained control over specific AI capabilities, to avoid the system impeding on deliberate diversions from writing norms: \textit{ ``Oftentimes in literary writing, we break grammatical conventions all the time... comma splices have become much more common in fiction writing, just because people use comma splices in real life all the time. ''} For W7, intentional rule-breaking was a distinctive aspect of their voice. Having the option to override AI suggestions that would “correct” these stylistic choices allowed them to preserve autonomy and authenticity in their work.
\end{enumerate}

These examples show how form-oriented writers prioritize their own stylistic and creative sensibilities, even in scenarios with established standards or grammar. They also point to the role that AI interaction design plays in supporting writers with monitoring and decision-making. W8 expands on this idea: \textit{I think for me, writing is so much about decision making that's like what you're doing at every single stage. And so I think that that's part of why I feel so attached to the AI being the one that's suggesting, but not necessarily the one that's  directly editing anything that you're working on...so it's important that those decisions are primarily made by you and not by the AI.''}

\section{Alignment Between the Two Studies}
\label{sec:alignment}
By comparing our literature review findings with our interview study results, we identify where existing design strategies address writers' concerns and where opportunities exist for more responsive system designs. This section analyzes alignment across three critical dimensions corresponding to the themes from \autoref{sec:study_2}: contextual factors affecting ownership concerns, writing process preferences across different contexts, and interaction design choices that shape writers' sense of agency.
\definecolor{ibmblue}{HTML}{648FFF}
\definecolor{ibmpink}{HTML}{DC267F}
\definecolor{ibmyellow}{HTML}{FFB000}

% Create light versions for table cells (20% opacity)
\colorlet{ibmbluelight}{ibmblue!20}
\colorlet{ibmpinklight}{ibmpink!30}
\colorlet{ibmyellowlight}{ibmyellow!30}

\begin{table}[ht]
    \centering
    \renewcommand{\arraystretch}{1.4}
    \resizebox{\textwidth}{!}{%
    \Large % Larger text size to counteract resizebox scaling
    \begin{tabular}{|c|c|p{3.5cm}|p{3.5cm}|p{3.5cm}|p{3.5cm}|p{3.5cm}|p{3.5cm}|}
        \hline
        \multicolumn{2}{|c|}{\multirow{2}{*}{\textbf{Cognitive Processes}}} & \multicolumn{2}{c|}{\textbf{Level of AI Support Demanded}} & \multicolumn{4}{c|}{\textbf{Level of AI Support Offered by Strategy}} \\ \cline{3-8}
        \multicolumn{2}{|c|}{} & \textbf{Content} & \textbf{Form} & \parbox[c][2.5em][c]{3cm}{\centering\textbf{S1: Structured Guidance}} & \parbox[c][2.5em][c]{3cm}{\centering\textbf{S2: Guided Exploration}} & \parbox[c][2.5em][c]{3cm}{\centering\textbf{S3: Active Co-Writing}} & \parbox[c][2.5em][c]{3cm}{\centering\textbf{S4: Critical Feedback}} \\ \hline\hline
        
        \multirow{3}{*}{\raisebox{-8\normalbaselineskip}[0pt][0pt]{\rotatebox{90}{\textbf{Planning}}}} 
        & \textit{Generating} & 
        \cellcolor{ibmpinklight} Strong user ownership over novel ideas with minimal AI input. & 
        \cellcolor{ibmyellowlight} AI-assisted ideation via prompts, but creativity retained by writer & 
        \cellcolor{ibmyellowlight} Ideas come from the user; AI helps them form connections and identify patterns & 
        \cellcolor{ibmbluelight} AI generates ideas which enumerate different approaches; user explores the idea space & 
        \cellcolor{ibmyellowlight} AI maintains the user's ideas while extending them or transforming them (e.g. keywords to prose) & 
        \cellcolor{ibmpinklight} Limited support for idea generation \\ \cline{2-8}
        
        & \textit{Organizing} & 
        \cellcolor{ibmyellowlight} AI supports in structuring ideas, retaining user ownership over logic flow & 
        \cellcolor{ibmbluelight} AI-assisted outline creation for enhanced cohesion in writer's tone & 
        \cellcolor{ibmyellowlight} AI helps the user to learn to structure their ideas in a particular domain & 
        \cellcolor{ibmyellowlight} User structures their ideas based on exploration through AI generations & 
        \cellcolor{ibmbluelight} AI assists with outline creation and structuring ideas & 
        \cellcolor{ibmyellowlight} Revision guidance supports organization of ideas following evaluation \\ \cline{2-8}
        
        & \textit{Goal-setting} & 
        \cellcolor{ibmyellowlight} Writer defines key objectives and frameworks; AI used for background structure alignment & 
        \cellcolor{ibmyellowlight} Creative goals developed by writer; AI supports structure adjustments & 
        \cellcolor{ibmbluelight} Scaffolding of AI system provides pre-defined objectives that must be followed by the user & 
        \cellcolor{ibmyellowlight} User defines goals, with AI assistance through iterative exploration and selection of ideas & 
        \cellcolor{ibmbluelight} AI works collaboratively towards writer-defined goals with some autonomy & 
        \cellcolor{ibmyellowlight} User maintains control over the text's goals; AI supports user goals through critical feedback \\ \hline
        
        \multicolumn{2}{|c|}{\raisebox{-5\normalbaselineskip}[0pt][0pt]{\rotatebox{90}{\textbf{Translating}}}} & 
        \cellcolor{ibmbluelight} AI used for drafting structured text; user ownership tied to novel ideas, not genre standards & 
        \cellcolor{ibmpinklight} AI assistance should be limited; primary voice retained through sentence-level decisions & 
        \cellcolor{ibmbluelight} AI content is integrated into the work through an iterative approach & 
        \cellcolor{ibmbluelight} AI provides both high-level (e.g. structural elements, plot) and low-level (e.g. dialogue) support & 
        \cellcolor{ibmbluelight} Users offload writing tasks to AI, emphasizing productivity and usability & 
        \cellcolor{ibmpinklight} Deliberate separation between AI and user workspaces, and manual integration of AI output limits translation support \\ \hline
        
        \multirow{2}{*}{\raisebox{-6\normalbaselineskip}[0pt][0pt]{\rotatebox{90}{\textbf{Reviewing}}}} 
        & \textit{Evaluating} & 
        \cellcolor{ibmbluelight} Grammar and clarity editing delegated to AI for efficiency & 
        \cellcolor{ibmpinklight} Limited AI assistance; primary voice retained through sentence-level decisions & 
        \cellcolor{ibmbluelight} Scaffolded feedback enables AI to deliver comprehensive evaluations to users & 
        \cellcolor{ibmyellowlight} User evaluates writing by comparing it to other AI generations & 
        \cellcolor{ibmpinklight} Limited support for user's text evaluation; AI is focused on generating content & 
        \cellcolor{ibmbluelight} AI systems provide qualitative and/or quantitative feedback on a user's text \\ \cline{2-8}
        
        & \textit{Revising} & 
        \cellcolor{ibmbluelight} AI-intervention to refine writing clarity is welcome & 
        \cellcolor{ibmpinklight} Strong sense of novelty in breaking conventions for stylistic effect and unique voice & 
        \cellcolor{ibmbluelight} AI generates proposals to help the user refine their work as a skill-building tactic & 
        \cellcolor{ibmbluelight} AI provides text in the user's workspace, enabling users to evaluate revised text in place & 
        \cellcolor{ibmbluelight} AI suggestions for revision are integrated directly into the text & 
        \cellcolor{ibmbluelight} AI offers fine-grained tools for specific revision tasks (e.g. summarizing) \\ \hline
    \end{tabular}
    }
    \footnotesize
    \textbf{Legend:} 
    \colorbox{ibmpinklight}{Pink - Little Support}; 
    \colorbox{ibmyellowlight}{Yellow - Moderate Support}; 
    \colorbox{ibmbluelight}{Blue - Significant Support}
    \caption{Comparison of AI Delegation Strategies Demanded by Study Participants and Offered by Strategies from HCI Literature, based on the support demands from participants in \autoref{sec:what_matters} and AI support from each design strategy enumerated in \autoref{sec: study1_strategies}. Cells are coloured by the degree of AI support demanded or provided, respectively.}
    \label{tab:combined-ai-support}

\end{table}

\subsection{Alignment with Contextual Factors of Ownership}
Writers' concerns about ownership in AI-assisted writing are contingent on specific contextual factors. Our interview study identified four factors that influence writers' concerns about ownership in AI-assisted writing: time constraints, level of trust, task importance, and perceived competence. These factors represent a user's personal value-based context or external limitations that shape their willingness to delegate writing tasks to AI. Our analysis indicates strong alignment between existing research priorities and writers' concerns. Researchers in HCI have worked extensively to investigate these dimensions, producing studies characterizing user values, social dynamics, and professional contexts \cite{gero23, biermann22, kim24, li24, rezwana23, inie23} that influence ownership preferences and how they shape users' attitudes toward AI assistance.

Several previous works within the CSCW community have addressed these factors within social and collaborative contexts. This body of work provides particular insight into how ownership concerns manifest when writing involves multiple stakeholders, shared authorship, or professional collaboration - contexts where questions of agency and ownership become especially complex. Time constraints, a key concern for our participants, have been explored by Shakeri et al. \cite{shakeri21}, who found that offloading narrative tasks to AI helped alleviate time pressures in collaborative writing, and by Cao et al. \cite{cao2023time}, who investigated how AI support systems can mitigate the effects of time pressure on decision-making. Trust-building in collaborative environments appears prominently in Hauptman et al. \cite{hauptman22components}, who found that professionals required explainable, actionable feedback and shared social context to build trust with AI collaborators, reflecting preferences similar to those of our interview participants. Perceived competence differences have been documented by Tang et al. \cite{tang24exploring}, who found differential usage patterns between professional and non-professional users of AI tools, and by Shakeri et al., who addressed users' vulnerability from lack of confidence in creative writing contexts. Task importance considerations appear in Zhang et al.'s \cite{zhang23framework} multi-level framework, which enables users to choose between different levels of AI creative involvement based on their individualized needs with a design that could accommodate varying levels of control desired for different stakes of writing tasks. While individual studies have addressed one or two of these factors, our analysis reveals how all four factors shape writers' ownership concerns in AI-assisted writing. This framework provides a foundation for future research to consider the full range of contextual influences on AI writing collaboration, rather than examining these factors in isolation.

\subsection{Alignment with Essential Cognitive Processes}
Our interview study revealed a critical distinction in what writers want to own, dividing writing contexts into two broad categories: Form-centric and Content-centric. As shown in \autoref{tab:combined-ai-support} no AI design strategy maps perfectly onto the delegation demanded by our participants. This highlights the importance of flexible systems that allow users to adjust AI involvement across different writing processes.

\subsubsection{Form-centric Writers}
The Creative, Personal, and General writing contexts afford writers greater freedom over form, allowing expressive personal styles. Form-centric contexts emphasize ownership over translation and revision while being more open to AI assistance with planning and ideation. As seen in \autoref{fig:table2-proportional}, these contexts had a mixed distribution of design strategies, with the plurality in each case being S3 (Active Co-writing). Since S3 prioritizes task efficiency and offloading work to the AI, this strategy may not fully address the needs of writers concerned primarily with Form contributions. For these writers all strategies offer more AI support in translating and reviewing than they demanded. We see awareness of this tension in systems that deploy S2 (Guided Exploration) methods of exploratory, iterative ideation which prompts creative writers to expand on ideas themselves. Research in this area, exemplified by \cite{schmitt21, gero19, kimmetaphorian23, difede22}, merits continued investigation to better support form-focused writers' sense of ownership.

\subsubsection{Content-centric Writers}
Content-centric writing contexts such as Academic and Formal writing prioritize communicating ideas with clarity and are subject to external stylistic constraints. Our interview participants in these contexts were primarily concerned with generating and organizing ideas and setting goals. For these writers, S1 (Structured Guidance) and S2 (Guided Exploration) are well aligned in terms of their Translation, Evaluation, and Revision AI support. As shown in \autoref{fig:table2-proportional}, S1 and S2 systems represented 61\% of systems in Academic contexts and 57\% in Formal contexts, demonstrating alignment between existing designs and the support demanded by our participants. Our analysis suggests these strategies offer more AI planning support than Content-focused writers desired. This indicates an area where users might benefit from proffering granular control over AI involvement. 

\subsection{Alignment with Desired Interfaces and Interactions}
\subsubsection{Suggestions}
Presenting AI content as suggestions is a common interaction design approach in AI writing systems, aligning well with users' demands. Researchers have investigated visual differentiation of suggestions \cite{osone21, singh23, bhat23}, enabling users to clearly distinguish between their own writing and AI-generated content. Other studies have examined the impact of suggestion length or quantity of suggestions on user experience and acceptance \cite{fu23, buschek21, dhillon24}, finding that suggestion length is inversely associated with perceived ownership of the text. The placement of suggestions within the interface also emerged as an important design consideration. Some systems present suggestions directly in the user's workspace \cite{chen19, buschek21, bhat23}, creating a more integrated experience but potentially blurring boundaries between user and AI contributions. More commonly, systems display suggestions in a separated interface \cite{goodman22, lehmann22, nichols20, dhillon24}. This separation creates a deliberate boundary that reinforces the writer's role as decision-maker, aligning with our interview participants' desire to maintain control over what enters their final text.

\subsubsection{Final Say}
Across the four design strategies we identified, each approach agency differently while supporting the principle of the writer having the Final Say. \textbf{Workspace Control} (S1) physically separates AI and user workspaces, ensuring changes require explicit user action. \textbf{Proposal Integration} (S2) presents AI-generated content as suggestions within an exploration framework. \textbf{Result Ownership} (S3) streamlines AI integration but potentially creates tension around authorship of the final product. \textbf{Analysis Separation} (S4) creates deliberate friction by requiring manual integration of AI-proposed revisions. Despite their differences, all approaches recognize that writers want to maintain editorial control. Across our dataset we did not encounter any systems that removed the writer's editorial control. However, some empirical studies \cite{draxler23} did have experimental conditions where the user had no influence over AI-generated text-which was associated with a reduction in perceived ownership.

\subsubsection{Global and Local AI Toggles}
We found that Global and Local AI Toggles are notably underrepresented in AI interaction research. While researchers such as \cite{draxler23, singh23, dhillon24} include control conditions with no AI assistance, our dataset contained no systems that offered participants the option of an AI toggle during normal operation. It was common that systems had user-initiated AI interactions, however this design choice does not fulfill our participants' desire for minimizing distractions or fine-grained control over how the AI interacts with their stylistic choices. This gap is noteworthy given that theoretical research on Human-AI collaboration frameworks, such as \cite{muller22, moruzzi24, shneiderman2022human} including CSCW research \cite{zhang23framework}, do investigate interfaces that modulate AI support as a mechanism for humans to exert control over AI initiative in complex tasks. The absence of these features in empirical design research presents an opportunity to investigate how toggles impact users' agency and ownership in practice. We encourage more research into systems where users can actively control their collaboration with AI and restrict assistance to designated components or remove it altogether.

\subsection{Monitoring}
Our analysis identified Monitoring as significantly underexplored in AI writing research. This high-level cognitive process becomes more complex with AI, as users must both monitor their own writing and oversee AI contributions. While monitoring as a cognitive process is distinct from the collaborative relationship between human and AI, they are connected through process management and a meta-level view of both the individual and collaborative writing processes. The gap likely stems from research focusing on optimizing specific interactions rather than examining broader collaborative dynamics. For instance, studies on suggestions do not allow participants to disable AI assistance entirely.

This represents a key research opportunity for CSCW. As AI systems advance, monitoring and management of the Human-AI collaborative relationship becomes increasingly important. The lack of research on monitoring and AI toggles suggests that current systems may not fully address writers' dynamic control over their collaboration. By developing more flexible interfaces that allow writers to modulate AI involvement, researchers could better support the nuanced relationship between assistance and ownership that emerged from our interview study.

\section{Discussion}

This paper, to our knowledge, is the first comprehensive study on designing for human agency within AI-assisted writing that combines a systematic review of generative AI-era research with an analysis of writers' perspectives on preserving agency and ownership. By considering both the state of the literature and user perspectives on ownership, we offer timely, actionable guidance to designers shaping the future of AI writing tools.

\subsection{Key Findings}

To structure our discussion of key findings, we divide the two research questions into four analytical components:

\begin{itemize}
    \item \textbf{RQ1.1 -- Design Strategies:} What design strategies are currently used or proposed in AI-assisted writing research, particularly in terms of interaction models and intended uses of AI-generated output?

    \item \textbf{RQ1.2 -- Process \& Context Distribution:} How are these design strategies distributed across different cognitive processes and writing contexts?

    \item \textbf{RQ2.1 -- Agency-Preserving Processes:} Which cognitive processes do writers consider essential to control in order to maintain their sense of agency during AI-assisted writing?

    \item \textbf{RQ2.2 -- Modulating Factors in Ownership:} How do user situations, writing contexts, and AI interaction types shape writers’ perceptions of ownership?
\end{itemize}

We answered the first research question encompassing components 1.1 and 1.2  through our systematic review and thematic analysis (\autoref{sec:study_1}).

\subsubsection{RQ 1.1 Design Strategies }

We identified \textbf{four primary strategies} for AI-assisted writing support that embody the existing literature: Structured Guidance (S1), Guided Exploration (S2), Active Co-Writing (S3), and Critical Feedback (S4):
\begin{itemize}
    \item \textbf{S1: Structured Guidance} supports users by providing step-by-step AI-assistance while fostering skill development (e.g., LitWeaver by Choe et al. \cite{choe24} leads novice researchers through completing a literature review)
    \item \textbf{S2: Guided Exploration} enables creative control through systematic exploration (e.g., ABScribe by Reza et al. \cite{reza24} enables users to rapidly iterate on chunks of text, storing previously-explored ideas and recipes for future exploration and revision)
    \item \textbf{S3 Active Co-Writing} facilitates efficient collaboration while maintaining user control (e.g. DiaryMate by Kim et al. \cite{kimdiarymate24} encourages users to select between AI suggestions to compose a diary that was meaningful to them)
    \item \textbf{S4: Critical Feedback} promotes strategies that facilitate user reflection and engagement through analysis and feedback (e.g. Impressona by Benharrak et al. \cite{benharrak24} specifies Personas that provide targeted feedback, prompting user reflection with a particular audience in mind)
\end{itemize}

\subsubsection{RQ 1.2 Process \& Context Distribution}

To identify how the design strategies are distributed across different cognitive processes and writing contexts, we mapped our dataset across five writing contexts adapted from Lee et al. \cite{lee24}---Academic, Creative, Formal, Personal, and General---and three cognitive processes based on Flower and Hayes \cite{hayes96}---Planning, Translating, Reviewing. This mapping revealed uneven research attention across processes and contexts (shown in \autoref{tab:table-1}), such as strong representation of AI support during the Generating subprocess within the Planning phase in Creative and Academic writing contexts, but limited focus on Monitoring support across all writing contexts.

Our investigation also revealed how these strategies are distributed across writing processes and contexts (see \autoref{fig:table2-proportional}), with S3 being the most widely used strategy overall (37.1\% of all systems in the dataset). Most notably, Creative writing showed a surprisingly high deployment of S3 systems (50\%), highlighting a potential misalignment between strategy use and user needs, as Active Co-Writing (S3) can compromise the sense of ownership central to writers' creative expression. We also found comparatively lower deployment of S4 systems (Critical Feedback), which accounted for only 14.5\% of the dataset. This uneven distribution suggests opportunities for expanding underused strategies like S4, particularly in writing contexts and processes where reflection and iteration play a central role.

While the strategies represent current research, they do not offer guidance on writers' values tied to preserving human agency. Our second study helps bridge this gap, and helps answer the second research question comprising components 2.1 and 2.2.  

\subsubsection{RQ 2.1 Agency-Preserving Processes:}

Our interview findings show that writers consider different cognitive processes as essential depending on what they see as their core creative contribution. \textit{Planning} is deemed critical by \textit{content-oriented writers} (e.g., academics and technical writers). These writers derive their ownership from generating original ideas, and prioritize information structure over style. Therefore, delegating this process to AI would undermine their sense of ownership. On the other hand, \textit{Translation} and \textit{Revision} are deemed essential by form focused-writers (e.g., poets and novelists), as their ownership is rooted in their voice, tone, and sentence or word-level  decisions. Maintaining control over these granular stylistic dimensions is central to preserving their identity as writers.

\subsubsection{RQ 2.2 Modulating Factors in Ownership:}

Writers’ perceptions of ownership are flexible, and are modulated by several factors, which we organize into three major themes spanning different \textit{user situations}, which influence \textit{when} ownership matters, \textit{writing contexts} influencing \textit{what} writers want to own , and \textit{AI interaction types}, which affect \textit{how} writers experience agency during AI-assisted writing. The first theme, covered in \autoref{sec:when_matters}, includes four key situational factors: time, importance, confidence, and trust. The second theme, covered in \autoref{sec:what_matters}, distinguishes between the two primary contribution types: content and form. The third theme, covered in \autoref{sec:interactions}, highlights four interaction types---suggestions, final say, global AI toggle, and local AI toggle---that modulate writers' sense of control when co-writing with AI. Together, these factors reveal that ownership in AI-assisted writing is not fixed but actively negotiated---shaped by writers’ goals, contexts, and the interaction design of the tools they use.

Each study offers useful insights on its own, but in combination they are more useful to designers because together they not only map the current research landscape, but also enable us to offer designers guidance on what \textit{should} be done to align with user demands, as explored in detail in \autoref{sec:alignment}. Study 1 is akin to a map handed to a sailor (the designer). Study 2 is akin to a compass that tells them where to go. Our findings indicate how writers’ sense of ownership is tied to specific cognitive processes: content-focused writers derive ownership primarily from ideation during the planning phase, as they feel that is where their primary contribution lies. In contrast, form-oriented writers connect their sense of ownership to translation and review, as that is where they want to exercise control over stylistic elements. This view of ownership suggests a ‘chessboard-like’ pattern, where users seek AI assistance in areas outside their primary contribution. \autoref{tab:categories} illustrates this preference: blue regions show where AI support is sought, pink areas denote places where AI should not intervene, and yellow regions denote zones where AI may assist with caution.

\subsection{Contributions to CSCW} 
Our work speaks directly to CSCW’s growing interest in human–AI collaboration in creative and knowledge work. While CSCW has traditionally focused on cooperation and collaboration between people—with computers serving as mediating tools—recent advances in AI have shifted this dynamic. As AI systems increasingly take on semi-autonomous roles, interactions with them begin to mirror human collaboration, carrying with them the ambiguity, social nuance, and negotiation once exclusive to human-human cooperation. Crucially, these interactions introduce new concerns around agency and ownership that our community now need to grapple with.

Recent CSCW programs reflect this shift, with dedicated sessions on Human-AI Collaboration and AI and Trust at CSCW 2023 \cite{cscw2023program}, and AI in Creativity Flows and Future Dialogues on Personal AI Assistants at CSCW 2024 \cite{cscw2024program}. Our work aligns with this trajectory by examining how writers interact with AI across distinct cognitive processes and writing contexts. By foregrounding the demands and boundaries writers seek to maintain, our study contributes both theoretical insight and practical design implications to CSCW’s ongoing conversations about how to build sociotechnical systems that support collaborative work—not just between humans, but with machines that now shape the creative process in increasingly social ways.
We also contribute to prior HCI research on mapping the design space of AI-assisted writing, such as Lee et al.'s 2024 exploration \cite{lee24}, by adding granularity to the fields' understanding of how to design for human agency. By decomposing writing into its component cognitive processes and situating them in distinct writing contexts, we surface new nuances in how agency and ownership concerns play out at the process-level. For instance, our findings enrich existing work on authenticity and ownership in AI-assisted writing. Gero et al. \cite{gero23} found that while authenticity and ownership are related, they are not directly correlated--users may not perceive a system that mimics their style as inauthentic. Our studies complement this by revealing that for some writers, particularly form-oriented and expert writers (e.g., W12), AI mimicry of style can feel deeply invasive. As W12 shared, ``I would feel a little \textit{violated}. For me, personal style is so signature to who I am.''

This example illustrates the value of pairing systematic reviews with user studies that go deep into areas of interest and importance to the research community, such as our focus on preserving human agency. Within that context, our work relates to broader theories in Human-Centered AI, such as Ben Shneiderman’s HAI framework, which argues that automation and human control need not be at odds on a unidimensional spectrum \cite{shneiderman2022human}, like in the classic 1978 characterization of automation by Sheridan and Verplank \cite{sheridan1978}. Instead, Shneiderman posits a multidimensional perspective where automation and control can increase concurrently, which resonates with our optimistic vision for AI’s role in augmenting human agency. Like Shneiderman's multi-dimensional characterization of automation and human agency, our approach demonstrates the value of viewing creative tasks as a multi-dimensional. By breaking it down into distinct cognitive processes and  contexts, we move beyond a one-size-fits-all perspective and highlight specific \textit{context} $\times$ \textit{process} dimensions where designers should focus AI support. 

\subsection{Limitations and Future Work}
Our study has limitations that warrant careful consideration when interpreting the findings. Firstly, our findings are influenced by our choice of theoretical framework \cite{flower81}. While the framework is widely used in AI writing research (\cite{biermann22, lee24, rapp2015thesis}) and provided a valuable lens for this study, it may not fully describe Human-AI interaction in creative and professional writing. Exploring alternative or complementary frameworks in future work could yield richer interpretations and better address the collaborative Human-AI or author-reader dynamics. 

Secondly, our systematic review's focus on the ACM Digital Library, while methodologically justified, presents a limitation to the comprehensiveness of our findings. Although our preliminary analysis demonstrated that the ACM Digital Library contained a substantially higher concentration of relevant papers (11\%) compared to other databases (2-3\%), this focused approach excludes potentially-valuable insights published in other venues. The ACM's disciplinary focus may have oriented our findings toward certain perspectives in computing and human-computer interaction, underrepresenting interdisciplinary approaches or perspectives from adjacent fields. Future research could include additional digital libraries to develop a more comprehensive understanding of the literature landscape surrounding AI-assisted writing.

Thirdly, while our inclusion criteria were broad, allowing participants aged 18 and above, the second requirement that participants have \textit{some} prior experience using AI tools for writing inadvertently limited the age diversity in our sample, resulting in a maximum age of 34. This excludes valuable insights from older adults who are also impacted by AI. Additionally, our sample exhibited limited educational diversity, with all but one participant having attended university. This educational homogeneity may have influenced perspectives on AI-assisted writing, as individuals with higher educational backgrounds might approach AI tools differently than those with other experiences. Given that LLMs serve as writing tools for many people beyond academic contexts, future studies should prioritize more educationally- and age-diverse samples to capture how different populations perceive and interact with AI writing assistance and ownership. The gender composition of our sample could also be expanded to examine gender-specific perspectives. Furthermore, as our study only included participants familiar with AI, our findings are less applicable to writers with no prior familiarity.  Future research could investigate the initial reactions and adoption experiences of AI-naive writers, illuminating potential barriers to entry and differing perceptions of agency in AI-assisted writing. 

Finally, our interview recruitment via social media and email invitations, combined with the relatively small sample, limits the generalizability of our findings. Our convenience sampling method may have introduced selection bias by primarily reaching participants from certain networks and communities, potentially overlooking diverse perspectives from the broader population and failing to capture the full variety of writing contexts, particularly in fields like creative writing and professional communication, where there are many different forms. We partially accounted for this by being selective in our recruitment, aiming to include writers with varied experiences, but a larger sample of writers can help further deepen our understanding of AI’s role across varied writing contexts. Additionally, although our participants had experience with a variety of AI writing tools, all had used ChatGPT, with fewer using alternatives. This concentration of experience with conversational tools, particularly ChatGPT, may have influenced how participants conceptualized AI assistance and limited their understanding of the broader AI writing design space. A larger and more diverse sample of writers using a wider range of AI tools can help further deepen our understanding of AI's role across varied writing contexts.

\section{Conclusion}
Our systematic review of AI-assisted writing research, combined with interviews with writers, shows that preserving agency and ownership in human–AI collaboration requires a nuanced understanding of when and how users seek control across writing processes and contexts. We identified four design strategies in existing research—\textit{structured guidance}, \textit{guided exploration}, \textit{active co-writing}, and \textit{critical feedback}—and found that preferences for AI involvement vary significantly depending on the writing task. Content-focused writers (e.g., academics) emphasize control over planning and ideation, while form-focused writers (e.g., creatives) value ownership in translation and revision. Drawing on contextual factors such as time pressure, trust, task importance, and perceived competence, we provide design guidance for adaptive systems that preserve user agency. This includes preferring AI suggestions over direct edits, maintaining clear authorial boundaries, and offering global and local AI toggles for modulating AI involvement. By aligning system design with the real-world needs of writers, this work lays the foundation for human-centered AI writing tools that enable true co-writing, on human terms.
% \appendix
% % \section{Post-Survey Likert-Scale Items}
% % \label{appendix:likert}
% % % \input{figures/questions_appendix}
% \section{Writer Profiles}
% \input{figures/writer_experience}

% % \begin{figure*}
% %   \centering
% %   \includegraphics[width=\linewidth]{figures/likert_scale-cropped.pdf}
% %   \caption{Likert-Scale Statements on User Perceptions of Cognitive Processes during Writing}
% %   \Description{Likert-Scale Statements on User Perceptions of Cognitive Processes}
% %   \label{fig:bar2}
% % \end{figure*}
\bibliographystyle{ACM-Reference-Format}
\bibliography{references}

\end{document}